\pdfoutput=1
\documentclass[namedreferences]{solarphysics}
\usepackage[hyperref,optionalrh,solaromanenum]{spr-sola-addons} 
\usepackage{graphicx}                    
\usepackage{color}                       
\usepackage{subfiles}                    
\usepackage{setspace}   
\usepackage{sidecap}    
\usepackage{subfig}
\usepackage{wrapfig}
\usepackage{caption}
\usepackage{booktabs}

\usepackage[pagewise]{lineno}

\usepackage[dvipsnames]{xcolor}
\definecolor{Mycolor}{HTML}{0E8A2B}
\definecolor{MycorrectionColor}{HTML}{FF0000}

\newcommand{\C}[1]{\textcolor{Mycolor}{\textbf{#1}}}

\renewcommand{\C}[1]{{#1}}

\edef\TeXjobname{\jobname} 
\edef\jobname{\detokenize{main}}

\begin{document}

\begin{article}

\begin{opening}

\title{\C{Matching temporal signatures of solar features to their corresponding solar wind outflows}}

%
\author[addressref={aff1},corref,email={diego.pablos.18@ucl.ac.uk}]{\inits{D.~de}\fnm{D.~de}~\lnm{Pablos}\orcid{https://orcid.org/0000-0002-5469-2378}}\sep

\author[addressref=aff1]{\inits{D.~M.~L.}\fnm{D.~M.}~\lnm{Long}\orcid{https://orcid.org/0000-0003-3137-0277}}\sep
\author[addressref=aff1]{\inits{C.~J.}\fnm{C.~J.}~\lnm{Owen}\orcid{https://orcid.org/0000-0002-5982-4667}}\sep
\author[addressref=aff1]{\inits{G.~V.}\fnm{G.}~\lnm{Valori}\orcid{https://orcid.org/0000-0001-7809-0067}}\sep
\author[addressref=aff2]{\inits{G.~N.}\fnm{G.}~\lnm{Nicolaou}\orcid{https://orcid.org/0000-0003-3623-4928}}\sep 
\author[addressref={aff3,aff4}]{\inits{L.~K.~H.}\fnm{L.~K.}~\lnm{Harra}\orcid{https://orcid.org/0000-0001-9457-6200}}

%
\runningauthor{de Pablos, Diego}
\runningtitle{Matching temporal signatures of solar features to their corresponding solar wind outflows}

\address[id=aff1]{Mullard Space Science Laboratory, University College London, Holmbury St. Mary, Surrey, RH5 6NT, UK}
\address[id=aff2]{Southwest Research Institute, San Antonio, TX 78238, USA}
\address[id=aff3]{PMOD/ WRC, Davos-Dorf, Davos, CH-7260, Switzerland}
\address[id=aff4]{ETH-Zürich, Hönggerberg campus, HIT building, Zürich, Switzerland}

\begin{abstract}


The role of small-scale coronal eruptive phenomena in the generation and heating of the solar wind remains an open question. Here, we investigate the role played by coronal jets in forming the solar wind by testing whether temporal variations associated with jetting in EUV intensity can be identified in the outflowing solar wind plasma. This type of comparison is challenging due to inherent differences between remote-sensing observations of the source and in situ observations of the outflowing plasma, as well as travel time and evolution of the solar wind throughout the heliosphere. To overcome these, we propose a novel algorithm combining signal filtering, two-step solar wind ballistic backmapping, window shifting, and Empirical Mode Decomposition. \C{We first validate the method using synthetic data, before applying it to measurements from the \textit{Solar Dynamics Observatory}, and \textit{Wind} spacecraft}. \C{The algorithm enables the direct comparison of remote sensing observations of eruptive phenomena in the corona to in situ measurements of solar wind parameters, among other potential uses.} \C{After application to these datasets, we} find several time windows where signatures of dynamics found in the corona are embedded in the solar wind stream, at a time significantly earlier than expected from simple ballistic backmapping\C{, with the best performing in situ parameter being the solar wind mass flux}.
\end{abstract}


\keywords{Solar wind, Origin $\cdot$ Heliosphere, Connection $\cdot$ Coronal Hole, Observations $\cdot$ Coronal Bright Point, Dynamics}

\end{opening}


\section{Introduction}\label{s:Intro} 
\label{S-Introduction}
Traditionally, the solar wind is categorized into different types depending on its plasma parameters as measured in situ. As the average solar wind speed is $500~kms^{-1}$, fast streams are classified as those moving at $\geq$ 600~$kms^{-1}$, and slow streams are said to be those moving at 450~$kms^{-1}$ or less \citep{Marion1980}. Fast, low density solar wind has been shown to be expelled from coronal holes \citep{McComas2008}, which are long-lived, predominantly monopolar regions of the solar corona that are expected to constantly release material into the heliosphere \citep{Krieger1973Apr}. The determination of the specific coronal origin of a given stream, as well as of the particular coronal regions which have important contributions is, however, more complicated. Several authors \citep[see e.g.,][and references therein]{Ohmi2004Jan, Sakao2007a, Stansby2019Jul, Rouillard2020} have suggested a number of methods that intend to aid in the determination of the specific origin of a solar wind stream, using the range of observations which are currently available. One of the fundamental reasons why the origin of the solar wind is hard to constrain is that a large amount of dynamic processing takes place during propagation, modifying parameters that otherwise might be linked to the specific near-Sun plasma properties of the stream \citep{Abbo2016Nov}. Fortunately, some solar wind plasma parameters are relatively unaffected by solar wind dynamical processes, such as the ionic composition of streams \citep{Bochsler1997}. Ion charge states become fixed in the solar corona, depending on the local electron density and temperature \citep{Wang2016}, and can therefore be used as a qualitative tracer for the likely origin of a given stream. In situ measurements of ion charge state ratios, along with calculations of the First Ionisation Potential (FIP) bias of the solar atmosphere through remote measurements, enable a direct comparison of ion populations of different elements that comprise the coronal plasma \citep{Pottasch1964, Stansby2020May}. While this can qualitatively help in determining the type of origin of a solar wind stream, it does not resolve the specific time and location of individual solar wind parcels. This is problematic for science which requires a verifiable connection between a solar wind source and its outgoing plasma.



One of the main science objectives of the recently launched \textit{Solar Orbiter} mission \citep[SolO;][]{Muller2013Sep} is to determine the origins of solar wind streams, shedding light on the origin and modulation of the heliospheric magnetic field and its charged particle population. For this reason, the spacecraft is equipped with a number of in situ detectors, as well as a range of imaging instruments. During close approaches at $\sim$ 0.3 AU, the high-resolution imagers will cover up to $\sim$ 6\% of the visible solar surface, and require a degree of targeting. The ideal pointing location for this science question is the source of the solar wind that is later measured in situ. Due to the delay between plasma ejection from the Sun and detection at the spacecraft location, modelling must be used to forecast likely coronal origins of a solar wind stream. The current methods estimating likely solar wind stream footpoints require knowledge of the solar photospheric magnetic field, modelling of solar wind propagation, and a range of observations of the stream properties. With regards to solar wind propagation models, some employ magneto-hydrodynamics (MHD) \citep[e.g.,][]{Mikic1999May} to resolve solar wind creation and evolution, and others assume simple ballistic propagation of the solar wind away from a rotating Sun \C{\citep[see e.g., the solar wind model by][]{Parker1958Nov}}. While the more rigorous MHD models are useful for generating a time-dependent, 3-D representation of solar wind flows, they take up a large amount of computational resources and time, and are unsuitable for quick predictions of stream footpoints for SolO remote observing operations. Simpler models that make use of the Parker-spiral-based ballistic back-mapping \citep[e.g.,][]{Nolte1973Nov}, paired with a potential coronal field model \C{\citep[PFSS;][]{Schatten1968}}, have been shown to perform almost as well as MHD simulations under simple coronal magnetic field configurations, such as those corresponding to simpler magnetic field configurations near the beginning of the solar cycle \citep[see e.g.,][]{0004-637X-653-2-1510, Allan2019Feb}. 

These simple empirical models are, however, more reliant on output verification to provide realistic information, as they are governed by over-simplified kinematics. To verify a given solution, different techniques have been suggested based on different solar wind coronal origin tracers \citep{Cranmer2019Jul}. These tracers have been taken to be compositional data (ionisation states and ratios), and structural coronal boundaries along with their effect on the outflowing solar wind. A potential solar wind origin tracer, which has not yet been explored with the aid of the relevant observations, is the potential inclusion of characteristic temporal signatures in the corona, and their propagation as part of the solar wind. \C{As these are numerous and highly variable, they could inject persistent temporal signatures in the outflowing solar wind. When these signatures are sufficiently different to other timescales observed in the corona, they could be used to aid in the determination of a solar wind source region for a specific stream.}

One example of these ubiquitous coronal dynamics are coronal jets. Solar coronal jets are highly collimated, transient structures observed in coronal plasma that are speculated to be produced due to magnetic reconnection between an underlying closed loop and an overlying open magnetic field \citep[see e.g.,][]{Crooker2012Nov, Edmondson2010May}. These jets last between 5 and 20 minutes, and due to their high occurrence rate across the solar atmosphere at any given time \citep{Raouafi2016Nov}, are thought to contribute a significant amount of mass and energy into the solar wind \citep{Kumar2019}. In remote sensing observations, coronal jets are linked to sudden localised enhancements of emission intensity in both the X-ray and EUV wavelengths that are coupled with collimated structures. Within in situ observations, rapid changes in radial magnetic field direction or \textit{switchbacks} have been observed in the solar wind at distances of up to 1 AU from the Sun \citep{Marsch1982Jan, Tenerani2020}\C{, with their occurrence rate being larger closer in to the Sun \citep{Kasper2019Dec}}. Several authors \citep[see e.g;][]{Owens2020, Sterling2020} conclude that a possible driver for these is plasma release from low-lying coronal loops through magnetic reconnection. This would make coronal jets that take place over open field (e.g a coronal bright point within a coronal hole) a potential source for these structures.

The \textit{Atmospheric Imaging Assembly} \citep[AIA;][]{Lemen2012Jan} onboard the \textit{Solar Dynamics Observatory} \citep[SDO;][]{pesnell2011solar} provides twelve second cadence, full-disk UV and EUV observations of the solar atmosphere at different temperatures. These observations have been successfully exploited to describe small-scale dynamics occurring in the solar atmosphere \citep[][and references therein]{Madjarska2019Mar}. By making use of the enhanced EUV emission in the SDO/AIA passbands, it is possible to characterise timescales and temporal behaviour of many of these events. We hypothesize that signatures of these dynamics may be captured in outflowing solar wind plasma parameters, which in turn present unique timescales often linked to solar coronal jets (5 to 20 minute sudden changes, with different timescales of variability found). 

\begin{figure}[!t] 
\centerline{\includegraphics[width=\textwidth]{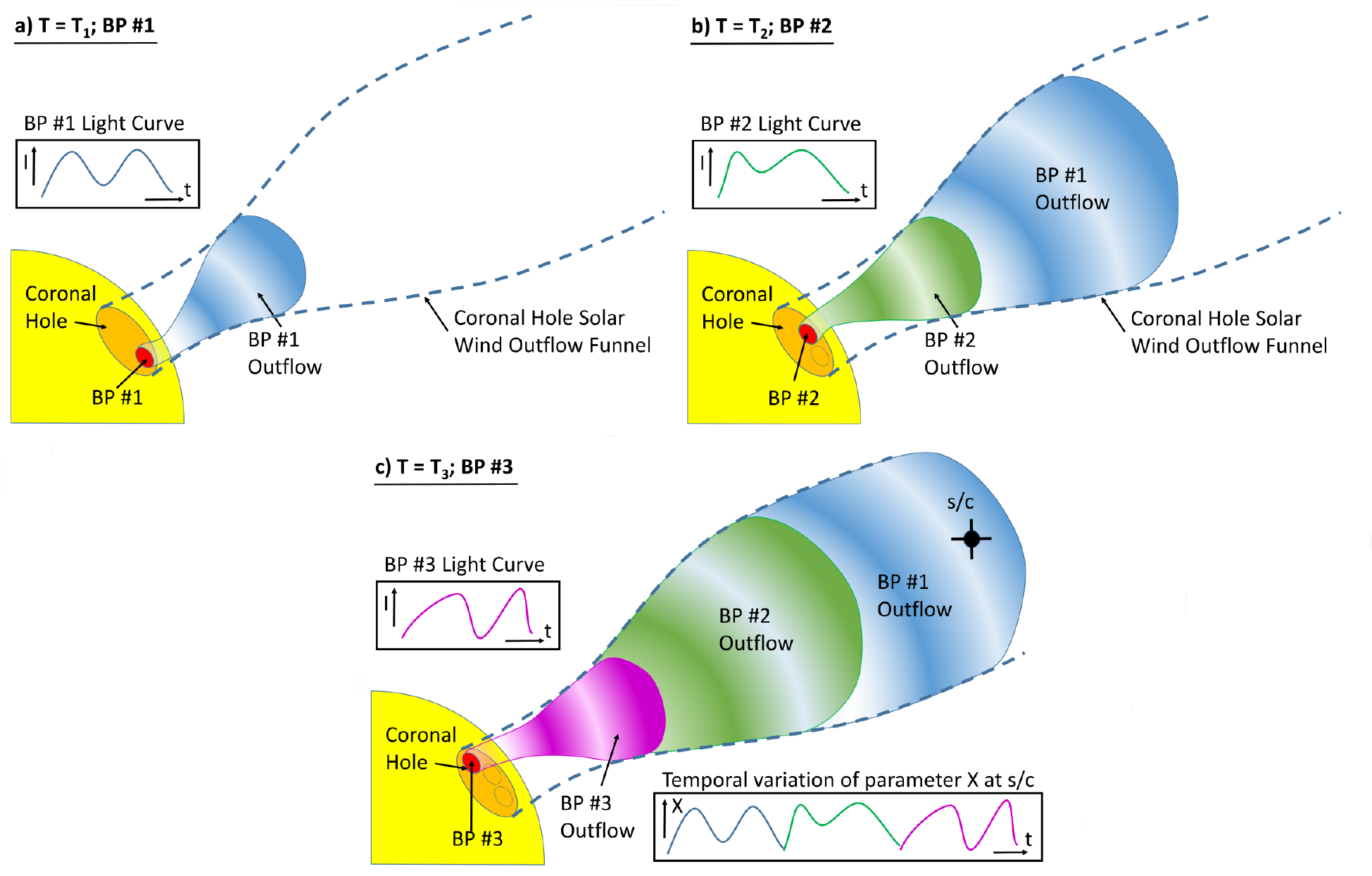}
          }
              \caption{\C{Schematic illustrating the hypothesis that the work in this paper ultimately seeks to provide a means to address. We argue that physical activity on the Sun, captured as signatures of temporal variations in remotely sensed measurements, may also embed similar temporal variations in some parameters associated with the outflow solar wind which is then sampled by a spacecraft. Here we illustrate the specific case of an Equatorial coronal hole exhibiting 3 successive bright points, identified by their light curves (illustrated on the left of each panel). Temporal variations either within the light curve of each bright point, or indeed across the events may relate to temporal variations in parameters describing the solar wind leaving the Sun. If the coronal hole is ballistically connected to a spacecraft via the associated outflow funnel, and even if the solar wind expands from the localised source into the outflow funnel, as shown in each panel, it is possible that the temporal variations from the source may persist and be identifiable at the spacecraft. There may thus be parameters within the solar wind flow which have a particular temporal fingerprint leaving the low solar corona and reaching a spacecraft in the form illustrated in the lower right of the third panel. If a match between the temporal variations of the light curves for a given bright-point, or for the coronal hole as a whole, and the observations at a spacecraft can be identified, this would provide evidence supporting the identification of the source region. In this paper we address a means to make this identification.}} 
\label{F-intro}
\end{figure}
Figure \ref{F-intro} shows three panels representing the evolution of \C{a solar wind stream that is ejected from an equatorial coronal hole and reaches a spacecraft (marked ``s/c" on the third panel). Within the coronal hole, a series of bright point eruptions take place, embedding their characteristic timescales, coloured in the same manner, into the solar wind. Each of the three panels is labelled by a distinct timestamp (T$_1$, T$_2$, T$_3$), for each bright point eruption and each resulting light curve (BP 1, BP 2, BP 3), such that each of the bright point eruptions embeds a specific signature on a solar wind parameter ``X". In this work we argue that because of radial propagation of the solar wind, for a given solar wind stream with an estimated source point, it is possible to compare dynamics observed at the approximate magnetic foot points to temporal variations found within solar wind stream parameters measured at the spacecraft location. Note that Figure \ref{F-intro} depicts magnetic field line expansion during solar wind propagation, which may have a significant effect on the signal morphology by implicitly stretching timescales or lowering amplitudes of variations and thus changing the signal over time, but these effects are not considered in this work. We instead concentrate on extracting a valid light curve containing signatures of coronal dynamics, and compare it against the different solar wind parameters.}

Observations extracted from coronal dynamics can however be expected to be highly nonlinear, not being fully described through a combination of strictly linear processes, as well as non-stationary, as their statistical properties can be expected to change over time. More specifically, in the remote sensing observations, the occurrence of particular jetting events or brightenings cannot be estimated through purely linear means, and their statistical properties are expected to broadly differ depending on local activity. Equivalently, many different processes may take place in the solar wind, increasing the complexity of the timeseries that is later observed in situ. Examples of such processes are plasma oscillations, turbulence, and solar wind inter-stream dynamics \citep{Velli1993Sep}. Due to this complexity, it is necessary to make use of a \C{signal decomposition} technique that allows for both nonlinearity and non-stationarity. Here we choose to employ Empirical Mode Decomposition \citep[EMD;][]{Huang1998, NewAs2011}\C{, which decomposes an original signal into a series of Intrinsic Mode Functions (IMFs), depending on timescales that are captured within them.} Applications of EMD in solar and plasma physics are varied; \citet{Long2017} utilise EMD in the characterisation of a trans-equatorial loop oscillation from remote sensing images. \citet{Linhua2013} and \citet{Kolotkov2017} employ the method to characterise long-term flare characteristics, and \citet{Stangalini2017May} use it for the study of MHD waves. 

In this paper, we present an algorithm that combines the extraction of a relevant signal from a coronal jet found within remote observations, linkage to in situ plasma measurements and windowing of the data for the application of EMD, allowing the extraction and comparison of temporal characteristics embedded within remote and in situ datasets linked to a coronal jet eruption. The algorithm is first tested for robustness by considering potential problems that come as part of the observations, such as noise or additional signal buildup, and is then applied to SDO/AIA and WIND observations of the solar corona and the solar wind, respectively.

\C{While we aim in future to use this technique on remote sensing and in situ observations made closer in to the Sun, the implementation shown here makes use of both remote and in situ observations taken from the solar corona and at the Lagrange L1 point, using light curves and in situ parameters respectively. Furthermore, the current approach is intended primarily for the verification of a potential connection, rather than the prediction of a best region to be observed.}

\section{Algorithm Overview}
\label{S-Algorithm}
\begin{figure}  
\centerline{\includegraphics[width=\textwidth,clip=]{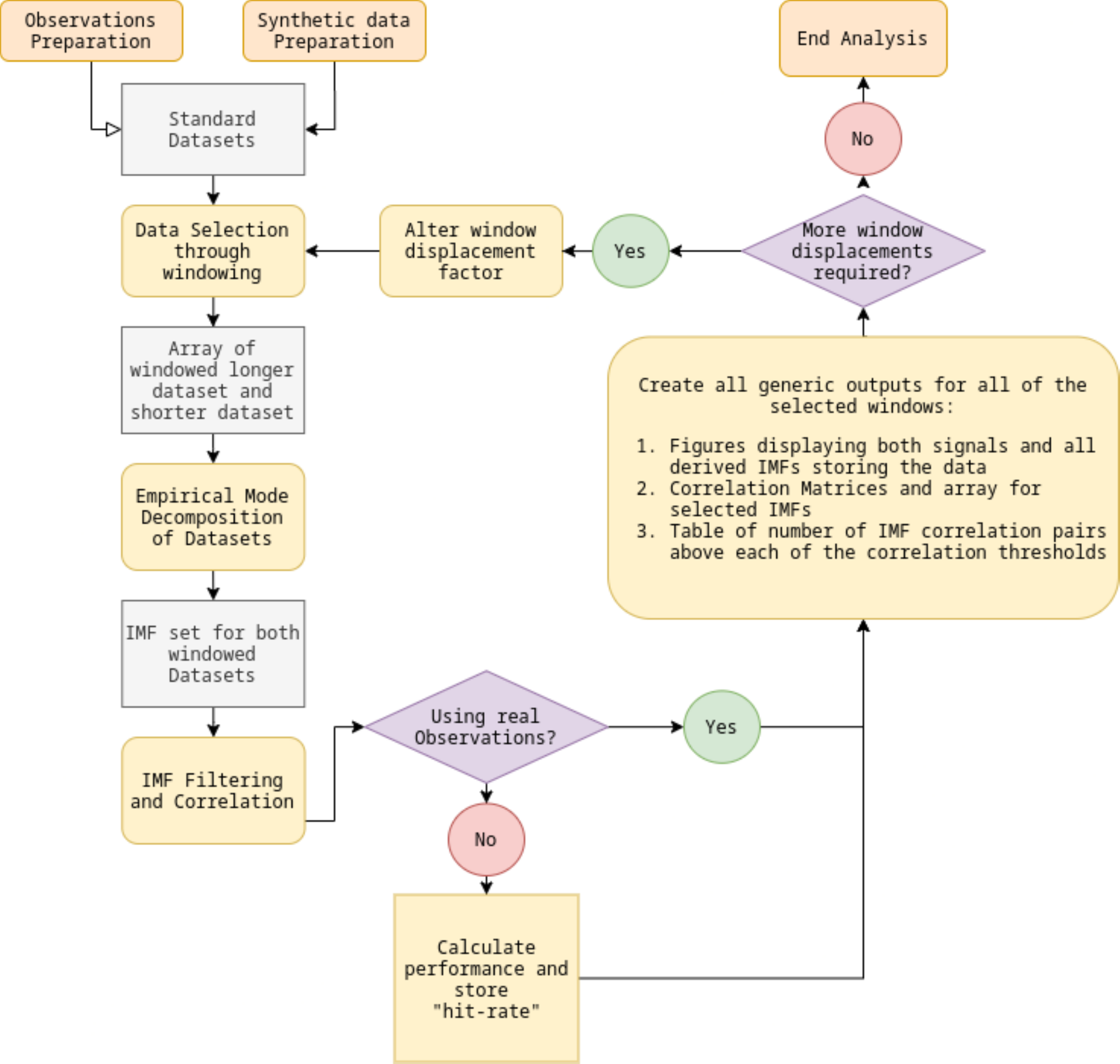}
          }
          \caption{Flowchart describing steps taken in timeseries analysis in this work. Briefly, the observations require preparation before starting the process, with the following steps describing an object (gray rectangles), and an action that is applied on it (yellow rounded rectangles). Details on each of the steps may be found in the text. \C{IMF is used as "Intrinsic Mode Function".}}
\label{F-breakdown}
\end{figure}

The main purpose of this algorithm is to find matching temporal signatures in two datasets of different duration. It achieves this through the use of a decomposition method that retains all the information in the signal through selective windowing of the longer of the two datasets. Figure \ref{F-breakdown} displays the steps taken from the input of a (real or synthetic) dataset to the conclusion of the algorithm analysis. To apply the algorithm, we begin by generating and preparing the datasets that are to be compared. We extract data from the long timeseries using a shifting window with the same duration as the short timeseries. For the short and each of the windows of the long dataset, a signal decomposition method is deployed to extract timescales of variability present in both in the form of functions, ignoring timescales not deemed relevant to the user. \C{We calculate the Pearson R correlation \citep{freedman2007statistics}, a measure of the linear correlation between two variables, for functions with relevant timescales, and generate a correlation matrix.} This procedure is repeated until the information included in the entirety of the long dataset is exhausted. Afterwards, the windows with statistically significant correlation identify the portions of the long timeseries where timescales of the short duration signal are present. 

\C{For a general application of the algorithm, there is an assumption that the two datasets have the potential for a physically meaningful correlation. Similar to extracting the linear correlation between two functions, it is important for users of the algorithm to ensure that the datasets they apply the method to can depend on one another either linearly or non-linearly.}

\subsection{Data standardisation, normalisation and windowing}
To prepare and standardise the data, we begin by ensuring that the two datasets \C{are continuous by linearly interpolating missing values. After this, if the sampling frequency of the datasets is different, we re-sample the higher cadence timeseries to the lower cadence through decimation.}

Once the datasets are on the same cadence and have been normalised, a window of the length of the short timeseries is used to extract windows from the long timeseries. This window is shifted by one, or any number of, time steps at a time, generating several windows of the long timeseries for later analysis. These windows are created until no more new windows of the same time duration as the short timeseries are possible.

\subsection{Empirical Mode Decomposition (EMD)}
\label{Ss-EMD}
When the data has been prepared and the relevant windows have been outlined for a specific pair of datasets, EMD can be used to determine the timescales of variability found in them. Appendix \ref{App:EMD} explains the application of EMD for any input signal in more detail. Briefly, the end result of applying EMD to a signal is a set of Intrinsic Mode Functions (IMFs), which are generated depending on characteristic time scales of oscillations present in the data. The first IMF captures the fastest timescales in the signal, and successive IMFs contain increasingly slower components. After all of the IMFs are successfully extracted, only the residual (trend) of the signal is left. Due to the approach taken by EMD, the original signal can be reconstructed exactly by taking the sum of all of its IMFs, and it is possible to make use of IMFs that capture specific timescales exclusively through consideration of their periodicity. EMD has furthermore been shown to outperform more traditional signal decomposition methods such as Wavelet or Fourier analysis when applied to datasets \citep{Huang1998}.

\subsection{Intrinsic Mode Function Selection and Correlation}
Since not all of the derived IMFs provide information that is relevant to the signal of interest, it is useful to consider only those that contain dynamics occurring within a range of periodicities. \C{P, the implied periodicity of an IMF, is calculated as;}

\begin{equation}
\label{Eq-period}
     P = 2T / n ,  
\end{equation}
where $T$ is the time duration of the IMF, and $n$ is the total number of maxima and minima in the IMF.

In our application to observations, we exclude IMFs with implied periodicities smaller than 5 minutes or larger than 100 minutes, as oscillations which are too fast are likely linked to background noise, and oscillations which are too slow and too similar to the trend may give oversimplified timeseries with respect to the dynamics of interest. After selecting the IMFs which are deemed as relevant for a given window, it is possible to correlate these against IMFs derived from the short dataset, effectively generating a correlation matrix similar to the one shown in Figure \ref{F-emd_matrix}.

\section{Algorithm Testing with Artificial Data} 
\label{S-Algorithm-testing}
When using synthetic data, it is possible to define the performance of the technique for typical scenarios. By generating a synthetic short dataset with a characteristic signature and embedding it within a long dataset, we can test whether high degrees of correlation in the IMFs of the short dataset and a window of the long dataset are found where the pulse is embedded, or if the high correlation was instead a false positive. 

We employ a correlation threshold corr$_{thr}$ to define a ``hit" as a window where temporal signatures of the two datasets match with an absolute correlation value stronger than the given corr$_{thr}$. Using this definition, a single window can display multiple hits in the case where several IMF pairs are correlated more strongly than corr$_{thr}$. By deciding how strict we are with IMF pair correlation, such that the number of false positives is minimised with respect to the number of correct identifications, we are able to optimise the performance for different use cases. 

For any given corr$_{thr}$, we may consider hits within the bounds of the real peak location to be $n_{correct}$, and total hits as $n_{total}$, declaring the hit rate as;

\begin{equation}
    \label{Eq-hrate}
    \textnormal{Hit~rate} = \frac{n_{correct}}{n_{total}} * 100.
\end{equation}

By using the hit rate, we may quantify how well the synthetic pulse location was predicted for a specific corr$_{th}$, taking values from 0-100\% depending on how well the real peak location was identified. 

\subsection{Results from artificial data tests}
For application and assessment of the performance of the algorithm under realistic constrains, artificial timeseries representative of expected coronal and solar wind properties were generated as follows. A \textit{short} timeseries (referred to as AIA$_{synth}$ hereafter) that displays properties of a typical solar signal of interest, i.e., an AIA spatially averaged intensity lightcurve of a jet observation, and a \textit{long} timeseries (WIND$_{synth}$ hereafter), within which the synthetic solar wind data was embedded, were created. We created AIA$_{synth}$ as a 12-second cadence timeseries with a duration of 1.5 hours, with a constant value of 40 arbitrary units. We then added a Gaussian pulse with a duration of \C{20 minutes (1200 seconds)} inside AIA$_{synth}$ to broadly replicate characteristics of a jetting region as observed by SDO/AIA. We created the WIND$_{synth}$ signal with a cadence of 3 seconds and a duration of 24 hours, taking a constant value of 10 arbitrary units, which are a similar cadence and average value for proton density measurements at 1 AU. We then decimated WIND$_{synth}$ to a cadence of 12 seconds, to match that of the AIA$_{synth}$ signal, and added the Gaussian pulse from within AIA$_{synth}$ to WIND$_{synth}$. 

After generating this simple short and long signal pair, for which we show results from the algorithm run on in Appendix \ref{App:algo}, different tests were used to evaluate the performance of the algorithm under some more realistic constraints. These were the addition of white noise to the datasets, and the addition of a secondary pulse within WIND$_{synth}$. 

\subsubsection{Effects of varying levels of white noise on signal recognition}
\label{Sss-NoiseInc}
By considering a range of white noise levels, we derive information about how well our technique would perform in a turbulent medium, as is the case for both the solar corona and the solar wind. Gaussian noise was added to both the AIA$_{synth}$ and WIND$_{synth}$ datasets, with a standard deviation of 1, 2.5, 5, and 10\% the amplitude of the embedded Gaussian pulse.
 
\begin{figure}[!t]
    \begin{center}
        \includegraphics[width=\textwidth]{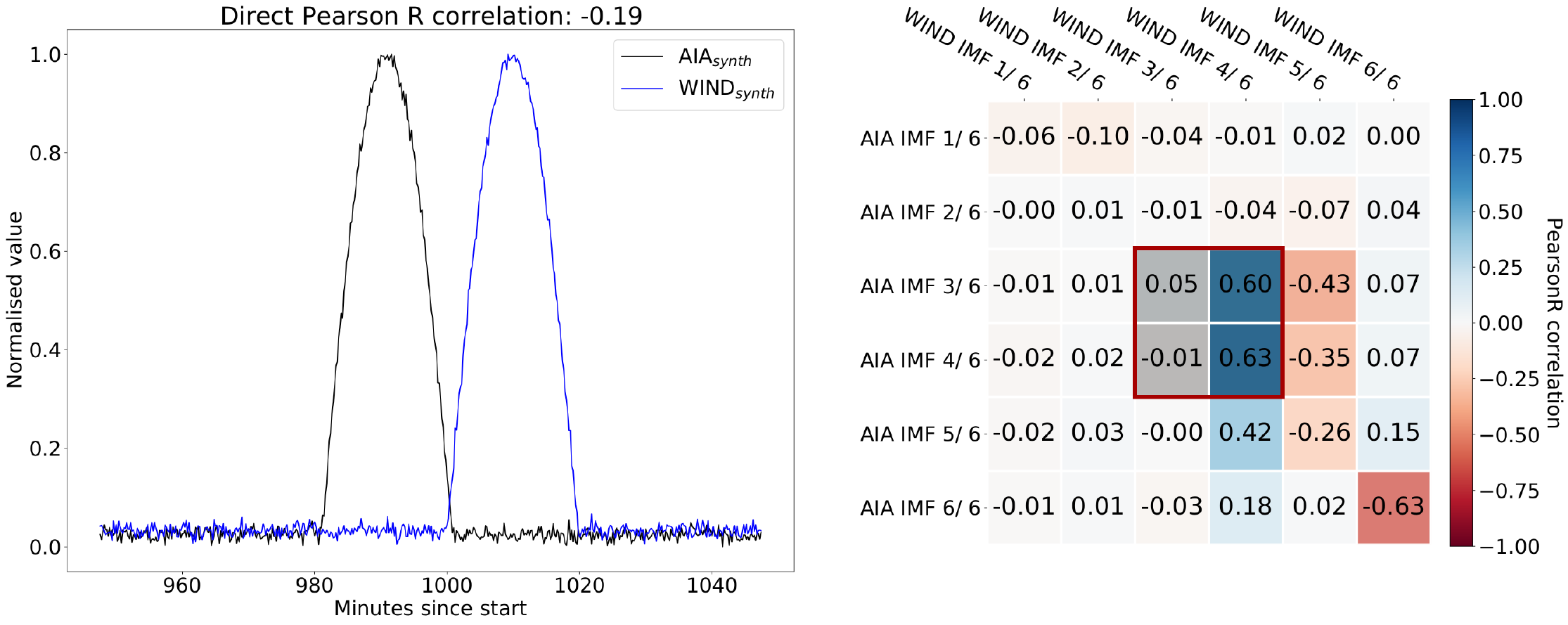}
    \end{center}
    \caption{Left: normalised value over time for a window of WIND$_{synth}$ (blue) and AIA$_{synth}$ (black), with white noise up to 1\% of the value of the peak, with the calculated linear correlation between the two on top. Right: Combined correlation matrix for the timeseries on the left, for all calculated IMFs, with colour linked with value of the linear correlation \C{shown on each IMF pair}. IMFs which are relevant for the analysis are highlighted with a red box.}
    \label{F-emd_matrix}
\end{figure}

In the left panel of Figure \ref{F-emd_matrix} we show AIA$_{synth}$ (black), and a window of WIND$_{synth}$ (blue), with their calculated Pearson R correlation of $\approx$ -0.19, where 1\% of the peak value is added as white noise to either signal. In the right panel of Figure \ref{F-emd_matrix} we show the corresponding correlation matrix for the derived IMFs for either dataset, with IMFs relevant for analysis highlighted with a red square. \C{Each of the squares is coloured depending on the absolute Pearson R correlation value for the specific IMF pair, and the numbers show the value rounded to two decimal places.} We observe that, out of the four IMF pairs that are \C{within the red square, and thus considered relevant}, two show a high degree of correlation ($>$0.6). We also note that, while directly calculating the Pearson R correlation between the two datasets leads to a very small linear anti-correlation of $\approx$ -0.19, the IMF matrix is effectively able to extract a stronger correlation, which may be expected of the two datasets by eye, as timescales included in both datasets are expected to behave similarly. \C{With regards to the significance of the results, we have added the two-tailed p-values correlations of another case study in Appendix \ref{App:pval}, using a similar number of measurements, such that the significance results are thus also relevant for these test cases.}

The correlation matrix \C{shown on the right side of Figure \ref{F-emd_matrix}} is useful to determine the likeness of the two \C{empirically decomposed} datasets for any specific window, but is lacking in terms of determining how well the signal was identified overall. For this reason, we examine the entirety of the WIND$_{synth}$ dataset, and identify all the windows where strong correlations were found.

\begin{figure}[!t]
    \begin{center}
        \includegraphics[width=\textwidth]{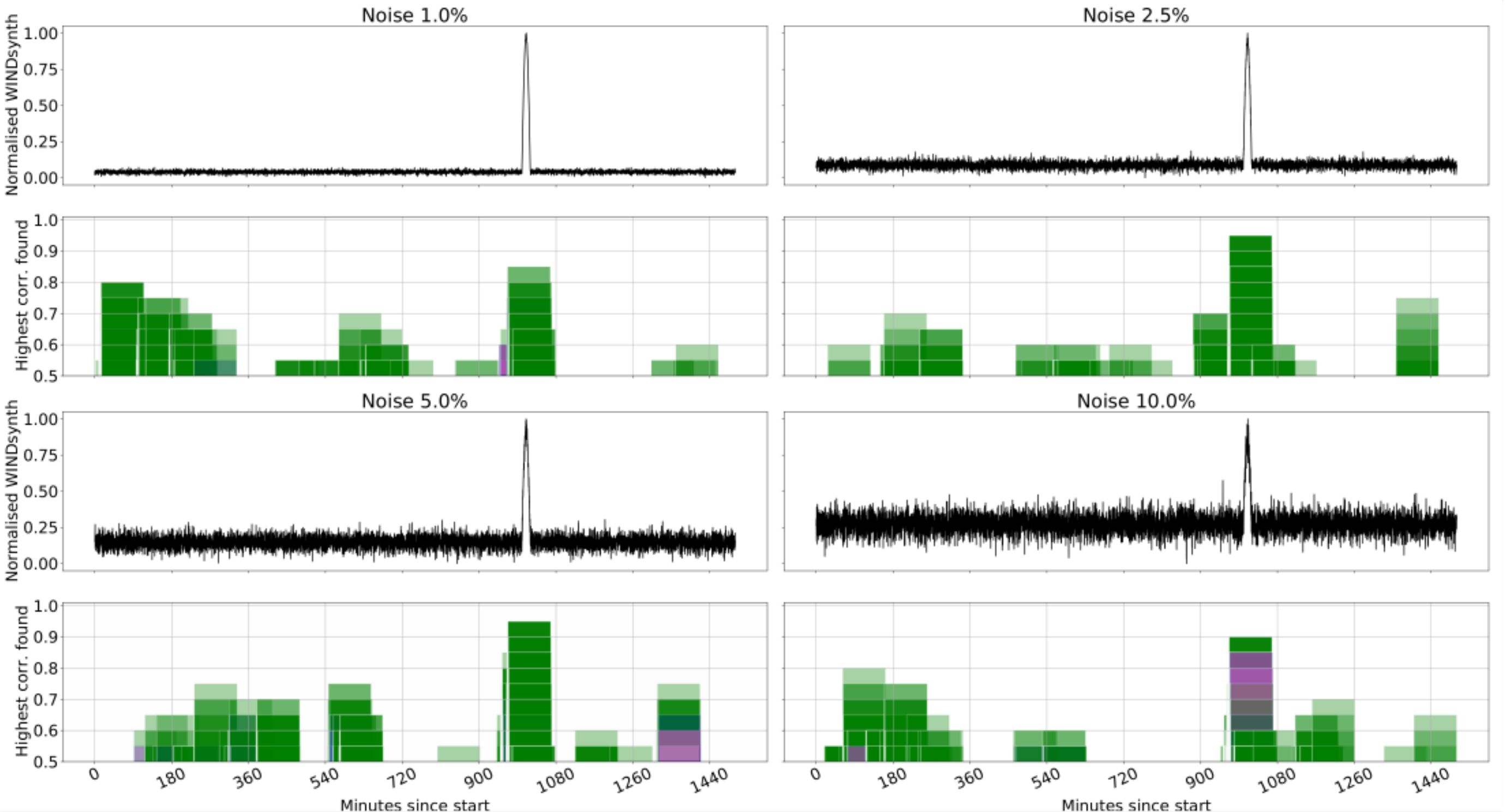}
    \end{center}
    \caption{Four panel figure, with each displaying normalised values of WIND$_{synth}$ (top), and windows (bottom) where considered IMF pair correlation is above a given corr$_{thr}$, for signals with up to 1, 2.5, 5 and 10\% of Gaussian noise compared to the amplitude of the embedded Gaussian pulse. For the bottom panels, the colour of the bar segments indicates how many IMFs display a correlation above the given level, with green for one pair, \C{and magenta for two}. Darker bar segments indicate an overlap of different windows, due to the small time shift between consecutive windows.}
    \label{F-noise_levels}
    
\end{figure}

Figure \ref{F-noise_levels} shows the number of IMF pairs (colour of bar in bottom plot for each) that reach a given correlation (Y-axis) for a given time window. We derive each of the panels in Figure \ref{F-noise_levels} from an AIA$_{synth}$, WIND$_{synth}$ signal pair with a given white noise percentage compared to the peak of 1, 2.5, 5, and 10\%. Within each of the panels, the top displays the normalised WIND$_{synth}$ signal over time, and the bottom displays the IMF correlation achieved at each of the derived windows. The width of the bar shown in the bottom was set to be equal to the time duration of AIA$_{synth}$, and the colour was set to green for cases where there was one hit (AIA-WIND IMF pair), \C{and magenta for two}. The opacity of the bar increases when hits are found consecutively. As we can see in Figure \ref{F-noise_levels}, most of the hits found at correlation thresholds higher than 0.85 are located in windows where the Gaussian pulse was embedded.

For summarised information about the results, Table \ref{T-noisethr} describes the hit rate for each of the noise levels, at each of the correlation threshold values. We see that 100\% hit rate is achieved for all noise profiles when the corr$_{thr}$ is taken to be at least 0.85, implying that selecting the window of any match above this corr$_{thr}$ will always yield a correct identification of the Gaussian pulse. On the other hand, we can see that more strict correlation thresholds of 0.9 or 0.95 lead to the cases with 1\% and 10\% noise respectively to fail to identify any IMF pairs as valid.

\begin{table}[!t]
\caption{Hit rate as calculated for a given Gaussian noise percentage at each of the correlation thresholds.}
    \begin{tabular}{l|lllllllllll}
        \toprule
          & 0.5 & 0.55 & 0.6 & 0.65 & 0.7 & 0.75 & 0.8 & 0.85 & 0.9 & 0.95 \\ 
        \bottomrule
        1\% & 14\% & 16\% & 22\% & 20\% & 24\% & 33\%& 40\% & 100\% & 0\% & 0\% \\
        2.5\% & 24\%  & 35\% & 51\% & 76\% & 84\% & 96\%& 100\% & 100\% & 100\% & 100\% \\
        5\% & 56\% & 60\% & 67\% & 74\% & 84\% & 93\%& 100\% & 100\% & 100\% & 100\% \\
        10\% & 61\% & 69\% & 80\% & 90\% & 90\% & 94\%& 97\% & 100\% & 100\% & 0\% \\
        \bottomrule 
        \label{T-noisethr}
    \end{tabular}
\end{table}

\subsubsection{Effects of additional signals on signal recognition}
\label{Sss-MultiSig}
Due to the large number of dynamic events taking place in the solar atmosphere, it is important to consider the effects of multiple pulses being captured in the in situ dataset. These pulses may add complexity to the original remote sensing signal, and could change the temporal signatures that are observed as part of the in situ data, changing the morphology of the derived IMFs. As the separation of signals depends on the timing of their release into the solar wind, it is necessary to explore the effect of an additional pulse on IMF pair correlation and hit rate with varying distances between the peaks. We choose to use a small amount of white noise of 1\%, similarly to the first test case in Section \ref{Sss-NoiseInc}.

To perform this test, the pulse found within AIA$_{synth}$ was isolated and duplicated, with the original and the copy being embedded in the WIND$_{synth}$ timeseries. The time difference between the peaks of the two signals was altered, selecting values from a quarter of the Gaussian width up to to three times the signal width, providing a range of WIND$_{synth}$ shapes. Extending the definition of hit rate from Eq. \ref{Eq-hrate} for this test, we considered hits as correct when either of or both of the Gaussian pulse peaks were included in the selected window.

\begin{figure}[!t]
    \begin{center}
        \includegraphics[width=\textwidth]{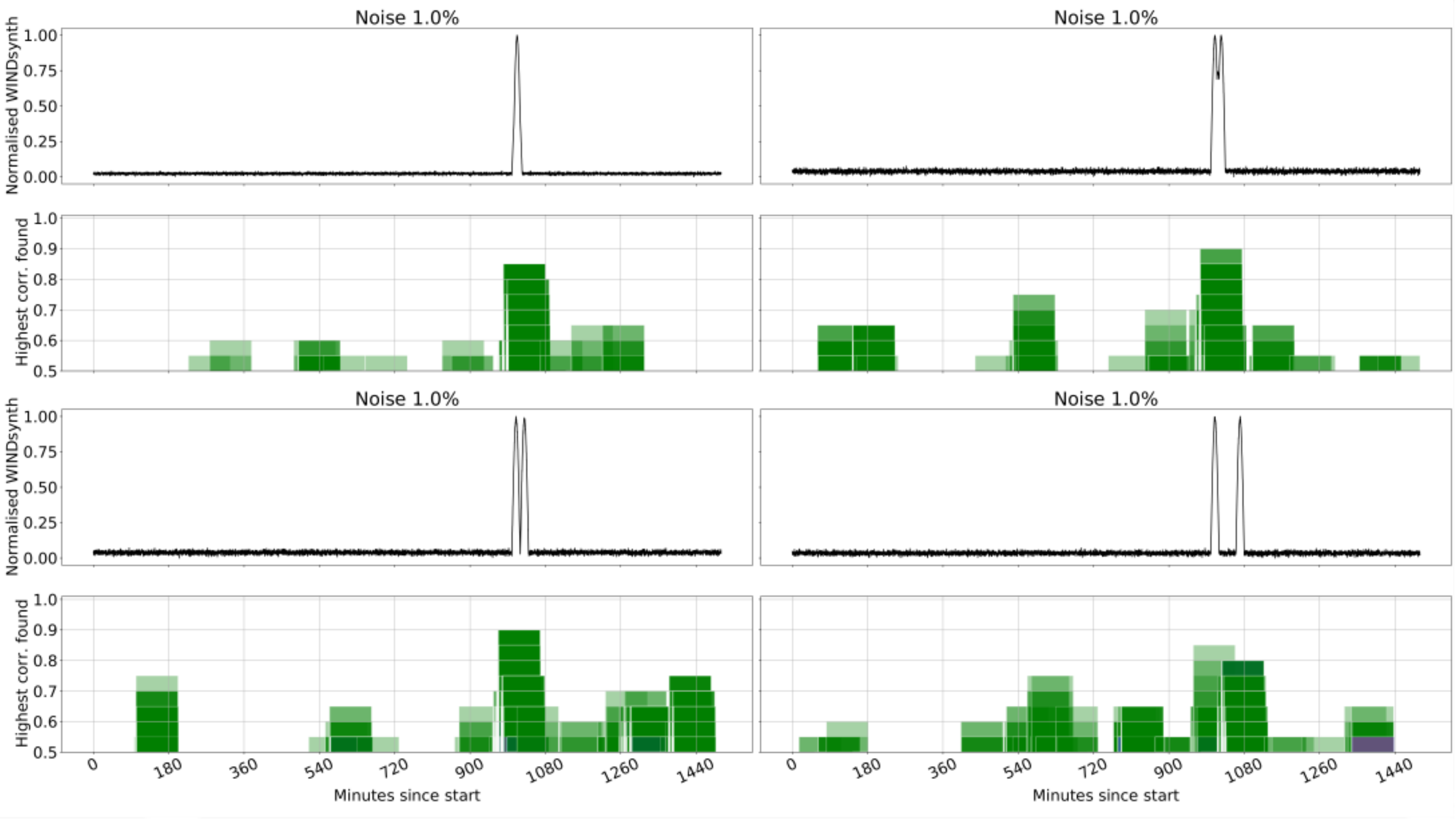}
    \end{center}
    \caption{Figure in the same format as Figure \ref{F-noise_levels}, with added white noise to AIA$_{synth}$ and WIND$_{synth}$ of 1\%, and a duplicated pulse in WIND$_{synth}$ positioned at 0.25, 0.75, 1, and 3 times the width of the original pulse from the centre of the original pulse.}
    \label{F-multi_signal}
\end{figure}

Figure \ref{F-multi_signal} shows the hit rate for a range of peak distances between the two signals on WIND$_{synth}$, with background noise taking values of up to 1\% of the peak signal value. Within each of the four test cases, the top panel shows the modified WIND$_{synth}$ signal, and the bottom shows windows where valid IMFs show correlations at each of the thresholds with the bars, with width equal to the duration of AIA$_{synth}$ and the colour of the segments being determined by the amount of short-long IMF pairs found to be correlated more strongly that each of the thresholds. The distances between the peaks of the duplicated signals tested here are taken to be, starting from the top left and following clockwise, 0.25, 0.75, 1, and 3 times the width of the original signal. All of the generated signals in this test can be directly compared to the performance of the first case on the previous test (top-left panel in Figure \ref{F-noise_levels}), as the white noise level is taken to be 1\% in both. We can see that most of the hits are found at the correct time if corr$_{thr}$ is set to 0.8.

Table \ref{T-multisig} displays the original results from 1\% noise on top of the simple signals on the first row, copied from Table \ref{T-noisethr}, followed by the hit rate results for the different peak separation distances shown in Figure \ref{F-multi_signal}. Table \ref{T-multisig} shows that, unlike in the previous 1\% noise test signal, a correlation threshold of 0.8 is enough for the hit rate of all cases to be 100\%, compared to the corr$_{thr}$ of 0.85 that is required on the case with 1\% noise. In Table \ref{T-multisig} we can also see that every one of the duplicated peak cases performs better than the simple 1\% noise test, with higher calculated hit rates across all correlation thresholds.

\begin{table}[!t]
\caption{Hit rate as calculated for the simple case with 1\% noise, and for each of the given duplicated signal peak separation, at each of the correlation thresholds.}
    \begin{tabular}{l|lllllllllll}
        \toprule
          & 0.5 & 0.55 & 0.6 & 0.65 & 0.7 & 0.75 & 0.8 & 0.85 & 0.9 & 0.95 \\ 
        \bottomrule 
        1\% & 14\% & 16\% & 22\% & 20\% & 24\% & 33\%& 40\% & 100\% & 0\% & 0\% \\
        \midrule
        0.25x & 57\% & 78\% & 86\% & 94\% & 100\% & 100\%& 100\% & 100\% & 0\% & 0\% \\
        0.75x & 38\%  & 47\% & 56\% & 64\% & 88\% & 91\%& 100\% & 100\% & 100\% & 0\% \\
        1x & 50\% & 53\% & 57\% & 61\% & 72\% & 80\%& 100\% & 100\% & 100\% & 0\% \\
        3x & 52\% & 64\% & 73\% & 80\% & 92\% & 92\%& 100\% & 100\% & 0\% & 0\% \\
        \bottomrule 
        \label{T-multisig}
    \end{tabular}
\end{table}

\subsection{Discussion of tests}
After assessing the performance of the technique under different scenarios, we have established the conditions under which the method performs poorly. Generally, a correlation threshold above 0.7 shows hit rates above 80\% in all cases, except the one with 1\% Gaussian noise. This is likely because too many IMFs are ignored in this case, as the signal is over-simplified with such a small amount of noise. Based on this assessment, we expect IMF Pearson R correlations above 0.7 to be relevant, and correct in three out of every four cases, therefore implying strong temporal signature correlation between the two datasets.

\section{Case Study of SDO and WIND datasets}
\label{S-CStudy_WIND}
    
    After the testing of the algorithm with synthetic datasets, the same methodology was employed to match temporal signatures of an observational case which used remote sensing and in situ measurements of solar features. In the case of coronal jet signals and the injection of their temporal signatures into the solar wind\C{, the assumptions we make for application of the algorithm are: 
\begin{enumerate}
    \item That dynamics observed as light emission intensity changes in the 193 Angstrom passband of SDO/AIA impose particular timescales that then propagate with the outflowing solar wind stream;
    \item That these timescales do not suffer significant changes during propagation out to 1 AU;
    \item That the Parker spiral model of the solar wind can be utilised to predict solar wind origins within a 12 hour error bar.
\end{enumerate}
}  

The first step to select data for a case study was to explore coronal hole solar wind streams near solar minimum, at a time where SDO observations were available. The chosen period was the second week of November 2016, where a coronal hole solar wind stream was identified through low charge state ratios, high bulk flow velocity, and low proton density in solar wind measurements at L1. This time period was checked using the \textit{Cactus Coronal Mass Ejection} catalogue (http://sidc.be/cactus/) to ensure there was no large scale, Earth-facing activity that may have affected the persistence of small scale coronal activity in the solar wind stream. In the remote observing side, a large southern coronal hole extension was identified during the relevant time period in SDO/AIA EUV images of the solar atmosphere, which spanned the solar equator and showed no active regions surrounding it. 

We first required a characterisation of observational signatures of relevant temporal signatures in the solar corona. Through investigation of these measurements, we inferred events that could potentially affect solar wind temporal characteristics. The measurements that we utilised for this purpose were light emission intensity measurements of the solar corona on the remote sensing side, and proton density, temperature, mass flux, solar wind velocity and magnetic field strength on the in situ side.

Due to the uncertainty in time of ejection and duration of travel, for the 1.5 hours of remote sensing observations, we collected 24 hours of in situ observations to apply the algorithm to.

\subsection{Observations}
\begin{figure}[!t]
\centerline{\includegraphics[width=\textwidth]{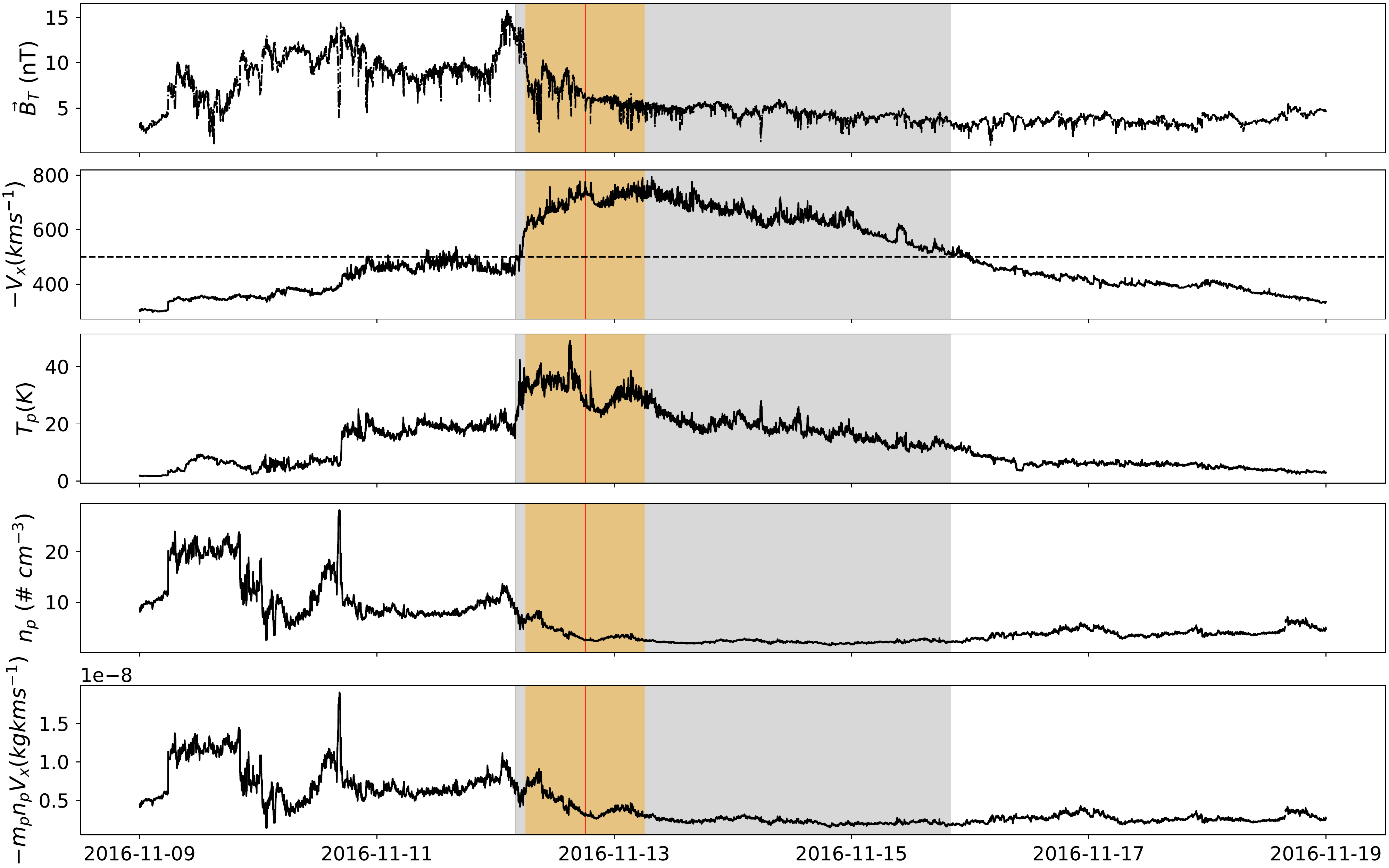}
          }
          \caption{Solar wind stream from 9 to 19 November 2016. From the top to the bottom panel, we show total magnetic field strength, solar wind radial velocity, proton temperature, number density, and mass flux. Highlighted in grey the portion of solar wind with coronal hole plasma-like properties: high bulk flow velocity (above 500 $km s^{-1}$) and low proton density. The red vertical line highlights data which is used for ballistic backmapping (18:05 - 18:15 12 November 2016). The orange column highlights data which is used in the analysis, selected up to 12 hours before and 12 hours after data used for backmapping.}
\label{F-sw_is}
\end{figure}

We use in situ solar wind measurements of plasma bulk flow velocity from the 3DP instrument \citep[3DP;][]{lin1995three} onboard the Wind spacecraft \citep[WIND;][]{harten1995design}, along with the spacecraft position at the time to perform ballistic backmapping of the solar wind stream to the solar corona. We highlight measurements used in the backmapping with the red bar in Figure \ref{F-sw_is} (18:05 to 18:15 UT on the 12 November), and the 24 hours of measurements used for analysis in orange. The backmapped stream shown in red was selected to be at the peak of outflowing velocity of a coronal hole solar wind stream, which lasts several days and is illustrated in gray. 

For the application of our algorithm, we used data from 06:00 UT on 12 November to 06:00 UT on 13 November 2016, and considered several different measurements from the WIND spacecraft. From the 3DP instrument, we considered the solar wind velocity $V_x$, proton number density $N_p$, proton temperature $T_p$, and calculated mass flux radially away from the Sun, $M_f= - N_p * m_p * V_x$. From the MFI instrument \citep{Lepping1995Feb} we calculated total magnetic field magnitude as $B_T = \sqrt{{B_X}^2 + {B_Y}^2+ {B_Z}^2}$.

To construct the remote-sensing signal, we employed EUV measurements of the 193~\AA\ passband from SDO/AIA. These provided us both with context information for the magnetic footpoint location, as well as with observations needed to characterise dynamics in the upper solar atmosphere. The coronal region of interest was selected using results from a combination of ballistic backmapping and coronal field line tracing, leading us to employ SDO/AIA 193 measurements from 15:00 to 16:35 UT on 9 November 2016. 

We obtained remote sensing measurements of the photospheric magnetic field as synoptic magnetograms. This is a data product that is published by the GONG network \citep[GONG;][]{harvey1996global}. The time of the selected GONG synoptic map was 18:00 UT on 9 November 2016, as this was the closest synoptic map to the calculated time of solar wind ejection of $\approx$16:00. These magnetograms contain measurements of the magnetic field strength and polarity, and were used as the boundary condition for a Potential Field Source Surface model of the coronal magnetic field. The PFSS model solver that we employed can be found in the ``pfss" package as part of SolarSoftWare, IDL \citep{pfss_ssw}. This package allows for the derivation of potential field solutions to the coronal magnetic field, with the free parameter being the ``Source Surface": defined as a height where the solar wind kinetic pressure is expected to dominate, such that any coronal loop which extends to the arbitrary height is considered ``open". The source surface height is usually set at around 2-2.5 Solar radii above the solar surface. In this analysis we selected a source surface height of 2.5 solar radii \C{to minimise the error associated with selecting a lower source surface and potentially backmapping to a region of the photosphere that has closed magnetic field.}

\begin{figure}[!t]
  \centering
  \subfloat[]{\includegraphics[width=0.8\textwidth]{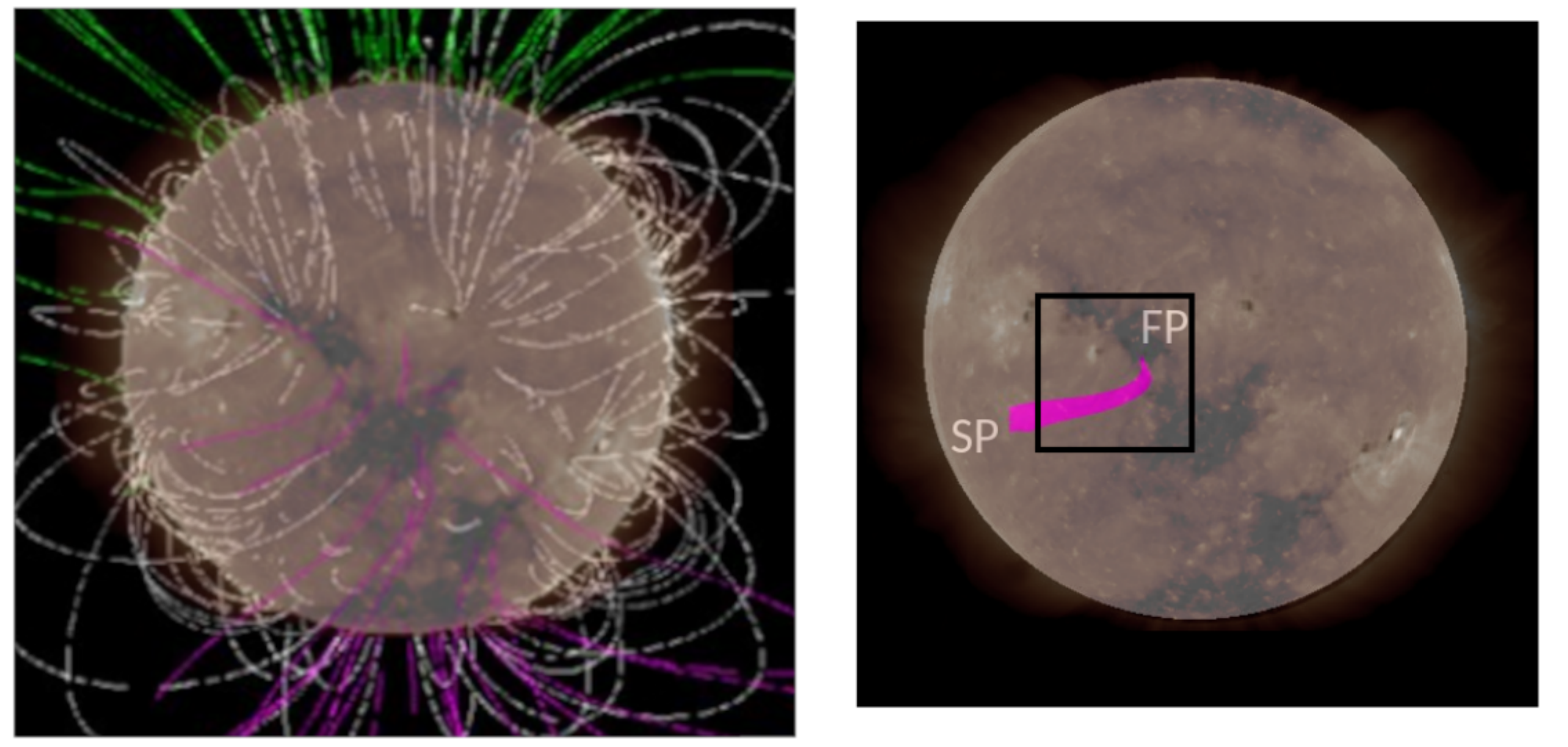}\label{Fs-pfss}}
  \hfill
  \subfloat[]{\includegraphics[width=0.8\textwidth]{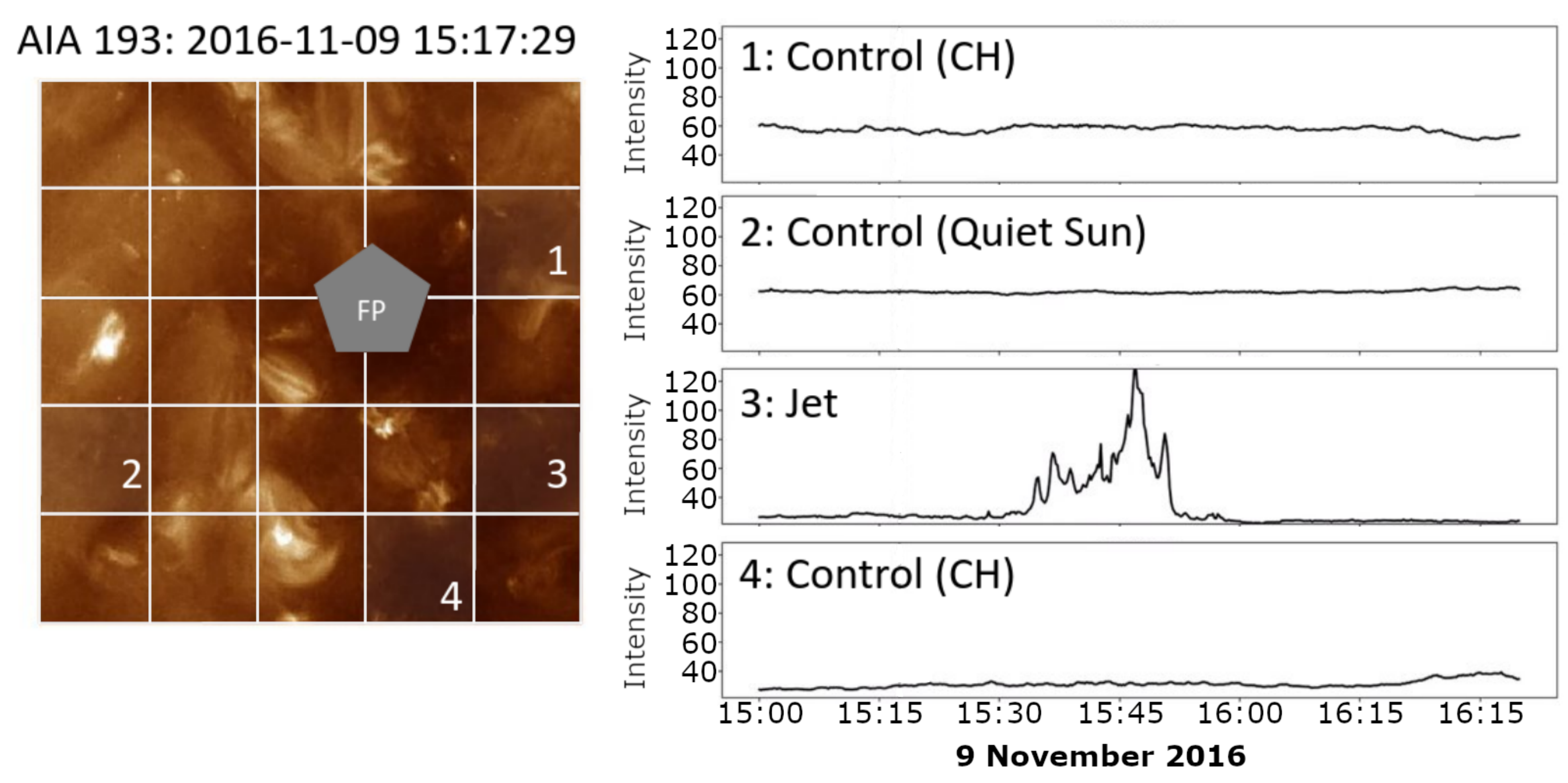}\label{Fs-aia}}
  \caption{a) Left: PFSS extrapolation overlaid on 193~\AA\ observation of the solar atmosphere, showing closed field lines in white, and open field lines in green (positive polarity) and pink (negative polarity). Right: Coronal field line tracing from source surface height of 2.5 solar radii, at the sourcepoint "SP", to the field footpoints in the solar photosphere, "FP", with the black bounding box representing the field of view used in the case study. b) Left: Cutout of an SDO/AIA 193 image of the area utilised in the case study, with numbered regions, and an indicated footpoint location for WIND observations (FP) with a grey pentagon. Right: Average emission intensity in SDO/AIA 193 for numbered regions outlined to the left, considering only values at least a standard deviation above regional mean for one hour and a half of observations.}
  \label{F-pfssaia}
\end{figure}

\subsection{Data driven modelling and Image processing}
\label{Ssec-model}

To link remote and in situ observations of the stream, we applied a two-step backmapping technique, commonly used to produce mapped solar wind sourcepoint locations at the photosphere \citep[see e.g.,][and references within]{Fazakerley2016}. This technique makes use of a  PFSS model, which combined with ballistic backmapping \citep[see][]{Nolte1973Nov}, allows for the establishment of a connection between a given solar wind stream, and a time and location of ejection from the solar atmosphere. We employed an empirical correction to travel time ($t_{real} = \frac{4}{3} * t_{ballistic}$), which was first discussed by \citet{Neugebauer1998Jul}, and later tested on coronal hole solar wind streams by \citet{Allan2019Feb}, showing promising results with regards to footpoint location when compared to the original calculated source region.  

Figure \ref{Fs-pfss} shows the PFSS extrapolation of the solar photospheric magnetic field at the relevant time, and overlays SDO/AIA 193 measurements to guide the reader in the relative position of open (closed) field lines coloured (white), on top of the equatorial coronal hole. On the right of Figure \ref{Fs-pfss}, a black square is overlaid on the figure, representing the location of the bounding box identified and used by the study, as well as calculated footpoints for the 10 minutes of WIND observations used in the ballistic backmapping, after following the relevant open magnetic field lines to the photosphere.

We took \C{these} magnetic footpoints to be within a $250$ arcsec$^2$ bounding box, from which a cutout of the selected pixels was selected to extract \C{the right panel of Figure \ref{Fs-aia}}. The size of this bounding box was chosen to be similar to the field of view of the High Resolution Imager that is part of EUI, onboard \textit{Solar Orbiter}, when the spacecraft is at 0.3 AU \citep{rochus2020extreme}. These Extreme Ultraviolet light emission intensity images were processed using the aia\textunderscore prep.pro routine, available in the SolarSoftWare library, before extraction of the lightcurves.

\C{From this $250$ arcsec$^2$ cutout we create a grid with a set of five rows and columns, separating each of the image into 25 squares, such that the small scale dynamics present within each of them could be well resolved by SDO/AIA, while some context was retained. Within each of these squares, we extracted a proxy for activity within the waveband as a lightcurve by considering the average of pixels which had a brightness at least one standard deviation above the regional mean.} \C{The right panel of Figure \ref{Fs-aia}} shows four EUV emission intensity curves (lightcurves) \C{for each of the numbered squares.} The third outlined region, ``Jet" shows data from a part of the coronal hole where a coronal bright point emerged and produced four brightening events which showed jet-like features (collimated intensity enhancements, fast flows), with durations over 10-20 minutes, lasting about 20 \C{minutes} overall. \C{The other numbered regions were extracted to be used as control datasets for both the signal extraction and analysis methods, and chosen to be different from the ``Jet" region in background intensity and expected timescales.}

We applied the algorithm to the four selected remote sensing datasets, as well as to each of the five parameters in the in situ dataset shown in Figure \ref{F-sw_is} for the period shown as shaded orange. The remote sensing data was taken to be the short dataset (AIA), and the in situ parameters were taken to be the long dataset (WIND). Similarly to the synthetic cases, each dataset was normalised before analysis.

\begin{figure}[!t] 
\centering
    \includegraphics[width=\textwidth]{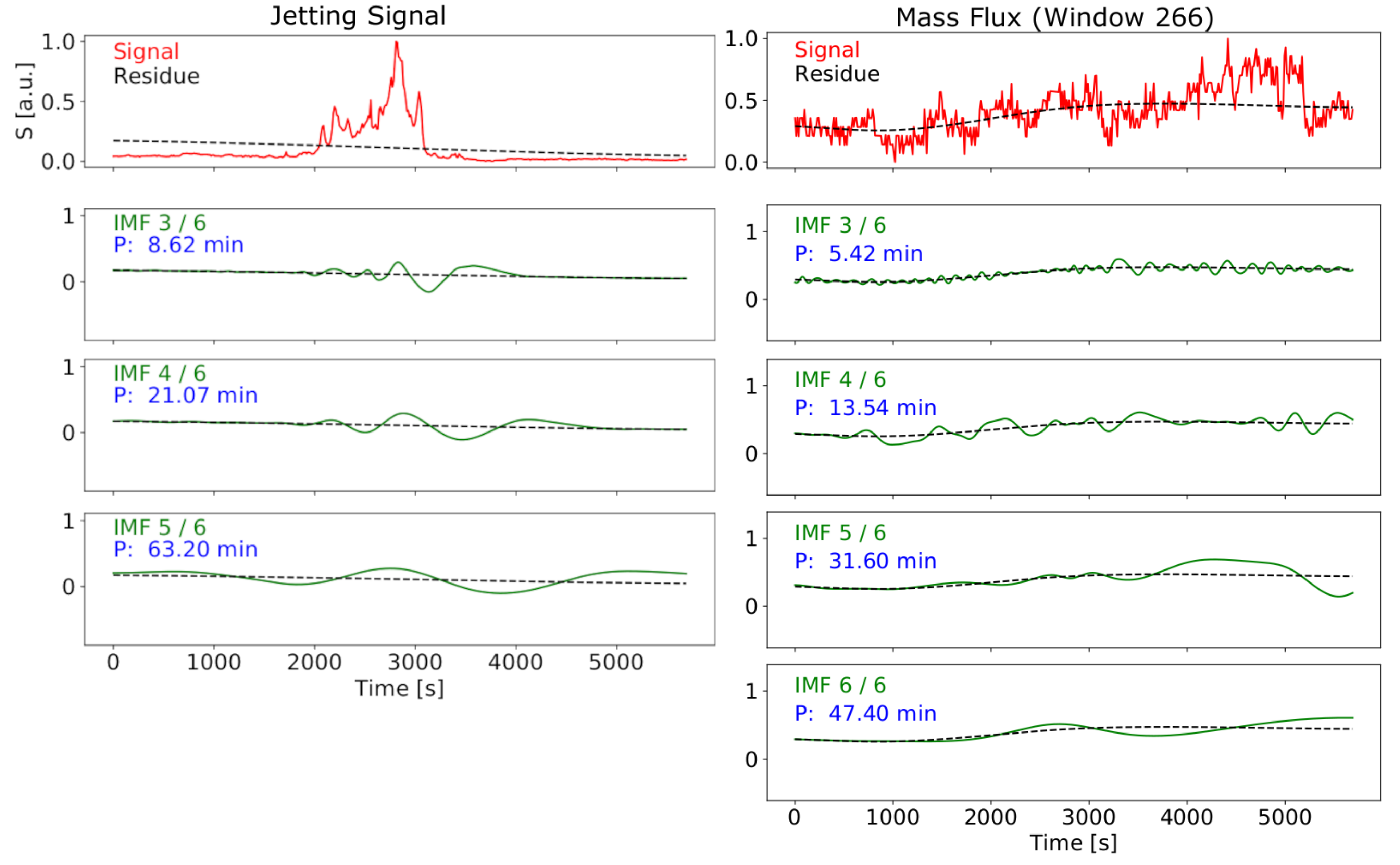}
      \caption{Left: Resulting IMF decomposition for normalised jetting signal of a region of the solar atmosphere that displays jetting activity on top of a coronal bright point. Right: Resulting IMF decomposition for a window of the proton density measurements which shows a high degree of correlation against the jetting signal. Top panel of each column shows original signal (red), along with residual (black), with panels below displaying all derived Intrinsic Mode Functions with the residual values added, from smallest to largest implied periodicities P, accompanied by IMF number.}
    \label{F-AIA_timespan}
\end{figure}

\subsection{Results}
Both of the columns in Figure \ref{F-AIA_timespan} display a signal decomposed using EMD, as well as their calculated IMFs with periods (P) found to be between 5 and 100 minutes, \C{as calculated using Eq. \ref{Eq-period}, and in line with timescales of coronal bright point jets \citep{Madjarska2019Mar},} along with the IMF number on the top left of each IMF. On the top panel of either column, we show the signal in red, and the residual in a dashed black line.

On the left panel of Figure \ref{F-AIA_timespan}, we show the ``jetting" signal. On the top panel we find a clear signal compared to the background, with a localised peak around the centre, and several features of varying magnitude within it, which are captured in the IMFs. IMF 3/6 captures dynamics that take place on approximately 10 minute timescales, capturing four sudden brightening events of different intensities. IMF 4/6 is constituted by dynamics that have a timescale of 20 minutes, namely the two main features of the jetting signal. IMF 5 captures the entire coronal bright point event, which shows an estimated period of 63 minutes, similar to the implied duration of the event. 

On the right column of Figure \ref{F-AIA_timespan}, we show a window of the mass flux signal that shows a high degree of correlation to the remote sensing dataset when applying the algorithm. Of all of the tested in situ parameters, the mass flux provides the highest correlation between the IMFs which are considered. On the top panel of Figure \ref{F-AIA_timespan}, we do not find a structure as clear as on the jetting signal. IMFs 3, 4, 5 and 6 show implied periodicities of 5, 12, 32, and 63 minutes, respectively, and are therefore all considered for correlation.

For each of the WIND parameters, the algorithm described in Section \ref{S-Algorithm} was applied, correlating them against the remote sensing jet observations, and thus yielding time shifts for which temporal signatures found within relevant IMFs up with those also found in the remote data. 

\begin{figure}
    \centering
    \includegraphics[width=\textwidth]{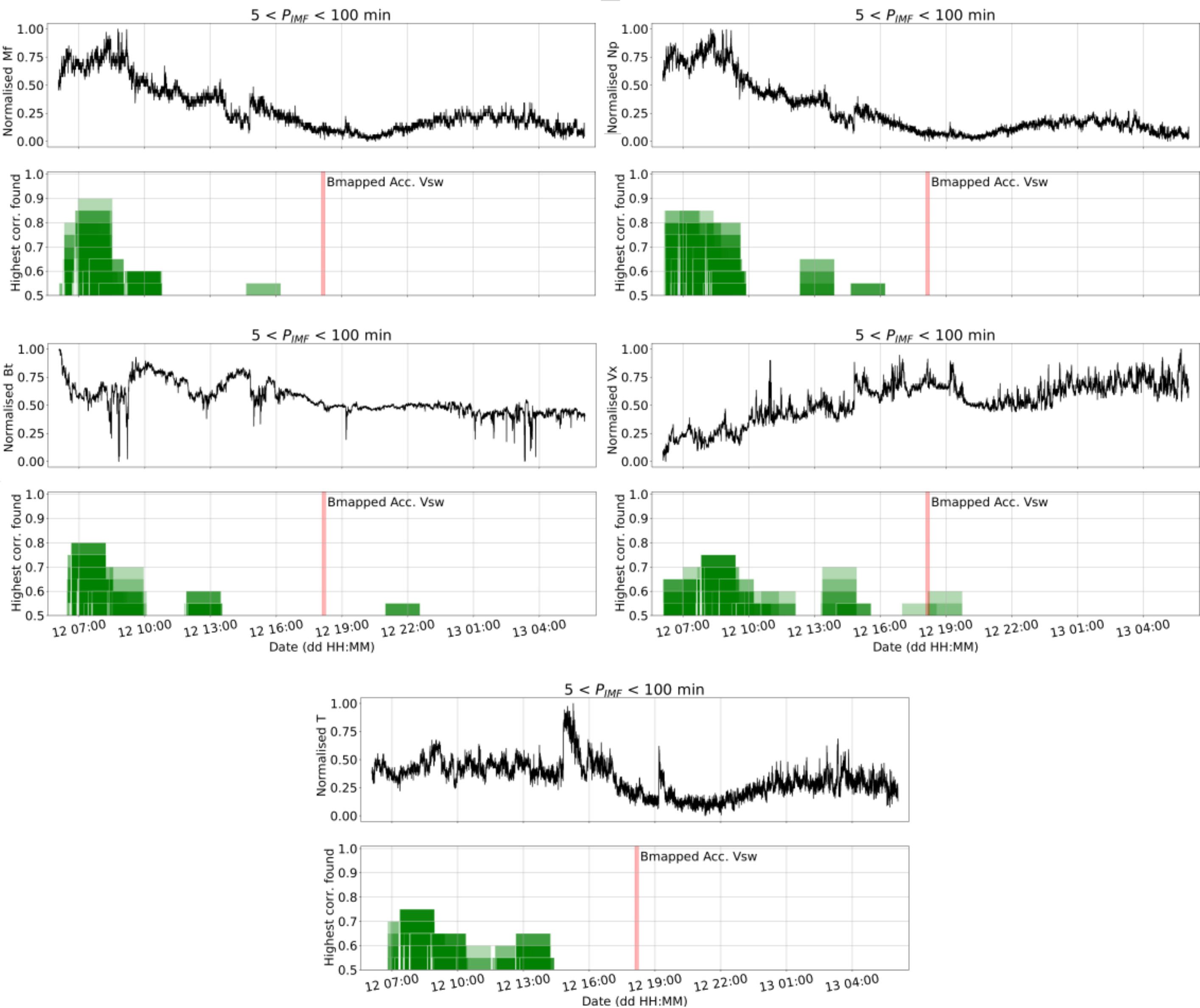}
    \caption{Figure format explained in Figure \ref{F-noise_levels}, for the set of WIND measurements used in the algorithm, decimated to 12 seconds and then normalised. The width of the green bars is equal to the time duration of the jetting dataset (1.5 hours), and the red bar shows the time used for ballistic backmapping. Variables shown are as follow: top left, solar wind mass flux (\C{$M_f$}) ; top right, proton number density ($Np$); bottom left, Total magnetic field strength \C{($B_T$)}; bottom right, solar wind radial velocity ($-Vx$); bottom centre: proton Temperature \C{({$T_p$})}.}
    \label{F-AIA_real}
\end{figure}

Figure \ref{F-AIA_real} shows the results from the algorithm being applied to \C{$M_f$} (top left), $N_p$ (top right), \C{$B_T$} (bottom left), $Vx$ (bottom right), and $T_p$ (centre bottom), ordered depending on the highest correlation achieved between valid IMFs. Within each of the five plots, the top part shows the normalised in situ dataset over time, and the bottom part shows the highest Pearson R correlation achieved for valid IMF pairs over time. The width of the bars is equal to the time duration of the remote sensing signal, and the colour of each of the segments corresponds to the number of IMF pairs which are correlated more significantly than each of the specific thresholds. The red vertical bar shown in all of the panels is co-temporal with the radial solar wind velocity used for pinpointing the origin of the solar wind stream.

Broadly, the highest correlations for all variables in Figure \ref{F-AIA_real} are reached near the start of the timeseries, approximately 10 to 11 hours before the backmapped time (07:00-UT on the 12 November). The variable that shows the strongest correlation is the mass flux \C{$M_f$}, with a window displaying a Pearson R correlation between one IMF pair higher than 0.9. The rest of the parameters shown here are correlated less significantly, displaying similar timescales to the jetting observations between 07:00-UT and 10:00-UT on the 12 November.

 \begin{figure}[t!]
\centerline{\includegraphics[width=0.69\textwidth]{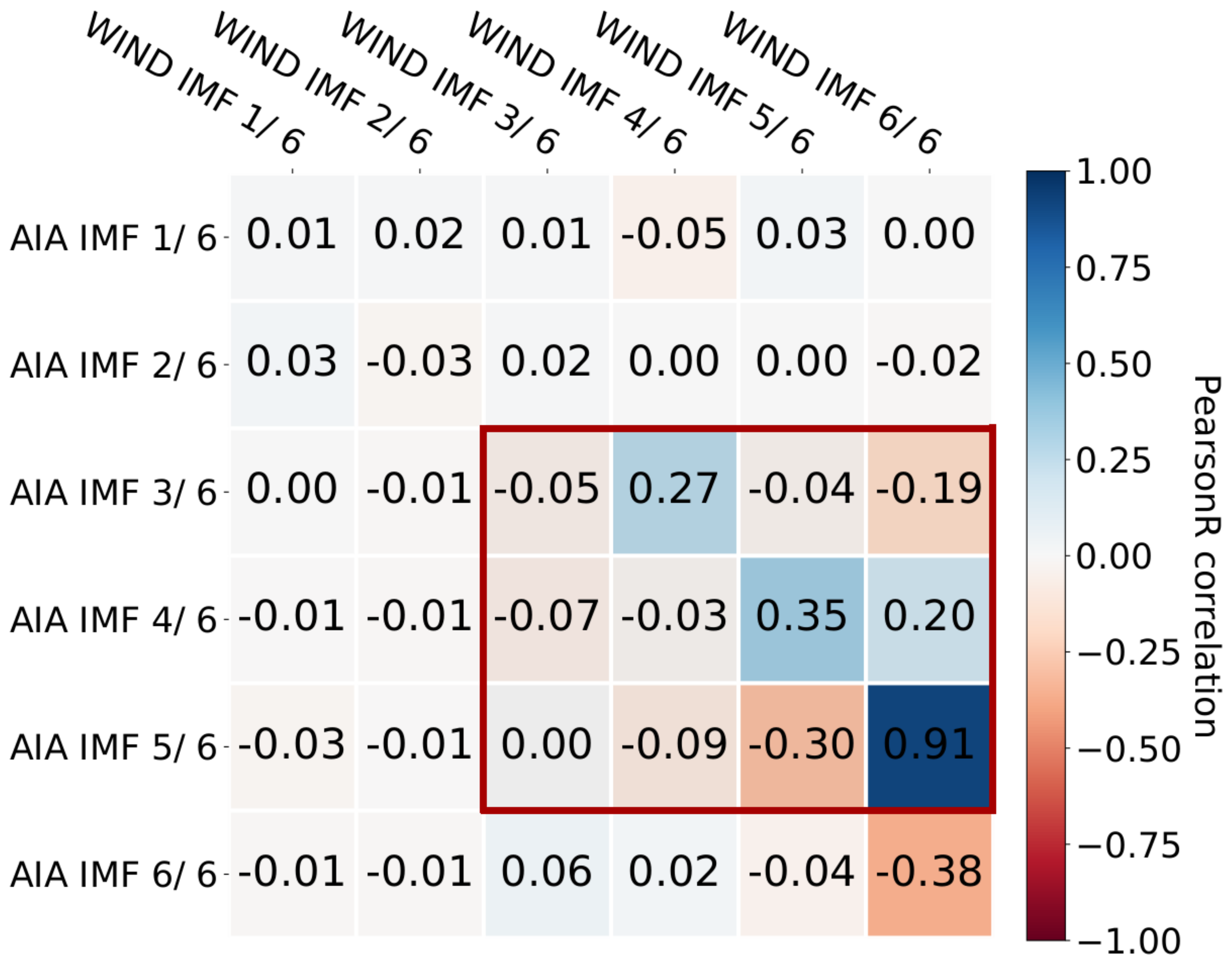}}
          \caption{Pearson linear correlation matrix for IMFs derived from a highly correlated window of the mass flux. Colour of the square indicates positive or negative correlation, with the intensity of the shade related to the absolute value. The highlighted rectangle in red, spanning WIND IMFs 3-6, and AIA IMFs 3-5, shows IMF pairs which were considered for correlation.}
\label{F-IMFcorrMatrixREAL}
\end{figure}
 Figure \ref{F-IMFcorrMatrixREAL} contains the IMF correlation matrix for a window of the normalised mass flux dataset that corresponds to the peak correlation reached on the top left panel of Figure \ref{F-AIA_real}. The intensity of the blue/red colour in Figure \ref{F-IMFcorrMatrixREAL} is linked to the strength of the calculated Pearson linear correlation between IMFs. The fifth AIA IMF is positively correlated with the sixth WIND IMF with a high value of 0.91. The remainder of the correlations highlighted in the red rectangle are below 0.4, and therefore are not strong enough to be relevant for the purpose of finding similar timescales within the two datasets. \C{Appendix \ref{App:pval} shows the corresponding p-values for the Pearson R values that we show in Figure \ref{F-IMFcorrMatrixREAL}, which are indicative of the significance of these results.}

\section{Discussion}
\label{S-Disc}
\subsection{Technique performance on artificial datasets}
We have described an algorithm that is able to correctly identify the location of a simple Gaussian-like pulse under idealised circumstances in synthetic observations. The performance appears to be poor when the utilised correlation threshold is below 0.7. We therefore see that the technique requires a strong, clear pulse, that needs to stand out well above background oscillations, and benefits from a degree of complexity on the background noise up to the case with 10\% noise.

When including an additional pulse in the dataset, the algorithm performance is better for larger separation between the pulses. This is likely due to a large overlap between the signals significantly deforming the original shape of the relevant window, affecting the derived IMF shape and by extension timescales which are detected.

\subsection{Technique performance on observational dataset}
\C{We have show that our} method can distinguish between coherent enhancements and random fluctuations of a signal by accounting for expected periodicity of events that occur in a remote sensing dataset (See Figure \ref{F-AIA_timespan}), while still considering the amplitude of the decomposed variability thanks to EMD. \C{For our case study, the correlation matrix for the window showing the highest correlation for the two datasets overall, the solar wind mass flux, is shown in Figure \ref{F-IMFcorrMatrixREAL}. We discuss the scientific implication for this result in the next subsection.} \C{We also note that the} only IMF pair which is highly correlated is that with the slowest variability for either dataset\C{, and see similar behaviour on} all of the other matching windows, where the highest order IMFs of either dataset are the ones found to be correlated more strongly. This is a different behaviour to that observed from the synthetic case, where a cluster containing several different IMF pairs shows strong correlation across them (see Figure \ref{F-emd_matrix}).

\subsection{\C{Science Results from observational case study}}

\C{After applying our algorithm to observations, we observe that the best match for the timescales of the jetting eruption occur in the solar wind dataset some 10-12 hours before the time identified using solar wind back-mapping. This high correlation is seen most notably in the mass flux but is identifiable also in the proton number density, total magnetic field, and across other analysed parameters with varying degrees of linear correlation values. Overall, these results imply that the solar wind from the coronal ejection could have reached 1 AU rather faster than would be expected from the simple ballistic back-mapping. The latter technique was applied in the case presented here with the ‘empirical correction factor’ of 4/3 originally suggested by \citet{Neugebauer1998Jul}. These authors employed this factor to force a closer match between mapping of coronal hole boundaries and their corresponding solar wind streams, and this factor has been subsequently used in other studies of solar wind propagation such as \citet{Allan2019Feb}. However, given this is a rather simplistic approach, it may be that one of the primary uses for our technique would be fine-tuning solar wind travel times for use in studies of the link between the solar corona and the heliosphere. Indeed, application of the technique for different events and over different radial distance ranges (either between the corona and spacecraft such as Solar Orbiter and Parker Solar Probe or between pairs of such spacecraft in the solar wind) offers the possibility of refining the value of the average empirical correction factor, and determining its dependence on distance range. In this sense, the technique developed here could offer the opportunity to determine the degree of acceleration the solar wind may experience at different distances. Nevertheless, in the case study presented here, our results imply a solar wind travel time of 64 hours instead of the 75 hours predicted by the simple back-mapping with a 4/3 correction factor.  Application of this factor implies that the solar wind has an average speed over its journey to 1 AU which is only 75\% of that observed. Performing the back-mapping without the 4/3 factor, the calculated travel time (for solar wind travelling at constant speed - the observed speed at 1 AU) would be 56 hours. Our result clearly falls between these two values, and so in this case we can conclude that the result would be consistent with the solar wind having an average speed which is 87.5\% of the observed speed at 1 AU. Since this speed is 17\% faster than use of the 4/3 empirical factor would imply, it appears that the solar wind stream had experienced less acceleration over its journey than the case studied by \citet{Neugebauer1998Jul}. This means that a smaller empirical correction factor of 64/56 = 8/7 would have performed better than the 4/3 acceleration factor for this solar wind stream.} 

\C{We performed the EMD analysis using a number of parameters categorising the solar wind. As we noted above, of these tested parameters, the solar wind mass flux is the parameter that shows the greatest similarity in EMD timescales with the coronal hole jetting signal. This may indicate that variation of the mass loss rate from the coronal hole, which manifests as variation in the mass flux detected at a spacecraft in the solar wind, has a broad relationship with the intensity variations in the bright point light curve (c.f. Figure 1).}  \C{The good correlation for the mass flux may then be due to this being a conserved quantity, decreasing at a rate directly proportional to radial distance.}

\subsection{Challenges in data collection}
\C{Several assumptions made when extracting the timeseries must be considered. These are with regards to the kinematics of plasma near and away from the solar origin, signal persistence and propagation, \C{impact of solar activity on performance,} or are imposed due to constraints of the modelling. }

\C{In terms of assumptions related to the kinematics of the plasma, there could be problems with the calculation of the backmapped time and location. The first potential problem is that the time taken for jet plasma to reach from the low corona to the point where the solar wind is generated, the ``source surface", is not considered here. Additionally, although we have argued above that the technique could be used to refine the empirical correction factor and more accurately identify a travel time, problems may result if in reality that correction factor, reflecting the degree of acceleration of the wind, changes rapidly from stream to stream.}

These issues can be minimised by making use of in situ measurements taken closer in to the corona, such as those provided by PSP and SolO, while the use of a different model for validation of the estimated emission time may help in terms of propagation within the low-lying solar atmosphere. 

Regarding assumptions related to the persistence of the signal during propagation, there is an important assumption that the morphology of the Intrinsic Mode Functions, and therefore of the maxima and minima present in the original signal and respective frequencies that are extracted, would be maintained from emission in the corona to the measurement at WIND. \C{Furthermore, the relationship between light emission intensity changes in the corona and plasma density and temperature has been shown not to be 1-1.} \C{In spite of the idea of modifying signal shape slightly being} discussed in the case where the duplicated signal overlaps with the original signal in Section \ref{Sss-MultiSig}, it could be further tackled by testing the effects of modifications of signal morphology through stretching \C{the signal} before insertion in the longer dataset, as well as through use of 1-dimensional solar wind models \citep[e.g.,][]{Pinto2017} to understand likely evolution and expansion of the solar wind and magnetic field, and thus of the signal.

\C{While we have tested the effects of one additional Gaussian pulse on synthetic observations, we have not considered the effects of a more active underlying corona. The current test case is purposely made on as quiet conditions as possible for available observations, and further work would need to be done in order to constrain and differentiate potential source regions of solar wind streams for more complex coronal configurations.}

Regarding assumptions made in the modelling of the coronal magnetic field, the selection of a representative source surface height for PFSS models has been studied extensively in the past. More recently, it was demonstrated that in order for the polarity of a coronal hole source region to match that of the solar wind captured at PSP, different modifications to the height of the source surface were required \citep[see][and references therein]{Badman2020}, with current work agreeing with a non-spherical source surface based on near-Sun observations from PSP \citep{Panasenco2020}. Similarly to the plasma kinematics problem, making use of PSP or SolO data to verify the connection through e.g. polarity on the two domains can help in deciding what the best performing source surface height is for the specific stream.

\subsection{\C{Implementation of the technique in Solar Orbiter operations}}
\C{
As it stands, the technique we propose has been tested to work on simple synthetic data, and on a single observational dataset. We show an approach for comparing timescales at a predicted solar wind stream footpoint to those present within the outgoing solar wind stream. For use in the Solar Orbiter data assimilation pipeline in the future, it will be necessary to investigate parameter space in more detail, and create a model for solar wind emission around the predicted footpoints, identifying likely areas, and contributions. Relevant parameters for remote measurements that can be refined are the selected fields of view, the size of sub-regions, and the magnetic field line expansion in the low corona leading to possibly different footpoints. With regards to in situ observations, the two free parameters are the variables being considered and whether there is a ``stretching" factor enforced by field line expansion to fill the heliosphere.}

\section{Conclusions}
\label{S-Conclusions}


We have created an algorithm that, through the use of windowing and EMD, is able to compare the timescales of variability of two different timeseries. We have tested the algorithm with synthetic datasets over a range of noise levels and found that, if using a correlation threshold of at least 0.7, we ensure temporal signatures are found where expected over 75\% of the time. We have investigated the effects of additional signals on the performance of the algorithm and have found that it was still able to extract relevant signatures with an even lower correlation threshold required. 

When we have applied the algorithm to SDO/AIA and WIND observations, we find similar variability timescales at a time ten to eleven hours earlier than expected on the basis of the solar wind stream used for backmapping, \C{and that the best performing parameter is the mass flux}. \C{The former finding implies} that the ballistic backmapping underestimates solar wind speed by about 15\% at 1 AU when using an empirical correction factor of 4/3 to simulate the effects of solar wind acceleration, \C{suggesting that a smaller empirical correction should be instead applied}. \C{We have argued that the latter finding is supported by the conservation of mass flux when assuming the magnetic field lines to be frozen into the plasma for high plasma beta environments.}

In the future, this technique is to be used with PSP and SolO data. To date, there have been no suitable remote observation configurations for PSP perihelia such that the footpoints of a stream are on-disk at the relevant time. The application of this technique to the closer-in PSP observations is therefore postponed for future work. 

Apart from making use of data from spacecraft found closer to the Sun, future work will explore signal shape modification through compressive effects in the solar wind by compressing and stretching the relevant remote sensing signal before applying EMD and comparing to the in situ observations. This approach would effectively increase the possible dynamical features that can be correlated, enhancing the applicability of this algorithm, as well as allowing for more realistic constraints. 

Additionally, and as a final planned application of the algorithm, we plan to apply it to in situ datasets found within the same Parker Spiral. With this application we will be able to verify its performance with equivalent properties for different types of solar wind stream, at varying distances to the Sun. As we have bulk flow velocity measurements of the solar wind at the spacecraft that is closer to the Sun of the two, ballistic backmapping will have a useful solar wind speed value. We will explore whether the times of statistically significant temporal signature correlation are when expected, or at a different time altogether. This will inform us on how under/overestimated the Parker spiral constant solar solar wind speed model velocity is, and by extension aid in the study of solar wind origins and propagation.

\begin{acks}
\C{The authors thank the referee whose comments helped to significantly improve the paper.}
This work utilises data from the National Solar Observatory Integrated Synoptic Program, which is operated by the Association of Universities for Research in Astronomy, under a cooperative agreement with the National Science Foundation and with additional financial support from the National Oceanic and Atmospheric Administration, the National Aeronautics and Space Administration, and the United States Air Force. The GONG network of instruments is hosted by the Big Bear Solar Observatory, High Altitude Observatory, Learmonth Solar Observatory, Udaipur Solar Observatory, Instituto de Astrofísica de Canarias, and Cerro Tololo Interamerican Observatory. DDP acknowledges use of NASA/GSFC’s Space Physics Data Facility’s CDAWeb service to access data (https://cdaweb.sci.gsfc.nasa.gov). The authors are grateful to the ACE, GONG, and AIA instrument teams for producing and making the data used in this study publicly available. DdP thanks the Science Technology and Facilities Council (STFC) for support via funding given in his PHD studentship. DML is grateful to the STFC for the award of an Ernest Rutherford Fellowship (ST/R003246/1). CJO and GN are grateful to the STFC for support via a consolidated grant to UCL/MSSL (ST/S000240/1). G.V. acknowledges the support from the European Union's Horizon 2020 research and innovation programme under grant agreement No 824135 and of the STFC grant number ST/T000317/1. This research made use of Astropy, a community-developed Python package for general Astronomy \citep{AstropyCollaboration2018Sep}, HelioPy, a Python package for space plasma physics data downloading and processing \citep{david_stansby_2020_4099097}, and of Sunpy, a community-developed Python package for heliophysics \citep{sunpy_community2020}.
\end{acks}

\clearpage

\appendix

\section{Empirical Mode Decomposition (EMD) definition}
\label{App:EMD}

The most evident advantage to using EMD over more traditional techniques is that, for any data, the sum of all IMFs plus residual is exactly equal to the input data, independently of data resolution and number of IMFs which are derived.

EMD analysis considers a given time series to be composed of a combination of oscillations which may not display purely sinusoidal behaviour. These oscillation packets can be isolated from the timeseries depending on their characteristic variability time scales. For an input signal $x(t)$, EMD($x(t)$) is calculated as follows;

\begin{enumerate}{\leftmargin=2em}

    \item Identify all local maxima and minima in the given timeseries (now taken as $x_{m}(t)$).  
    
    \item Interpolate between all identified local maxima, to estimate an upper envelope $x_{high_{m}}(t)$, and between all local minima to estimate a lower envelope $x_{low_{m}}(t)$.
    
    \item Compute the mean of the two envelopes: $m(t) = \frac{x_{high}(t) + x_{low}(t)}{2}$ and subtract it from the given timeseries: $d(t) = x_{m}(t) - m(t)$, then set $x_{m+1}(t) = d(t)$.
    
\end{enumerate}

Steps i-iii are then repeated on the newly defined $x_{m+1}(t)$ until the resulting signal, $x_0$ satisfies the following criteria:

\begin{itemize}

    \item That the number of extremes and the number of zero crossings are equal, or differ at most by one.
    
    \item That the mean value $m(t)$ of the upper and lower envelope is zero at all times.
    
    \item That the standard deviation difference between two consecutive repetitions is not smaller than a predetermined value.

\end{itemize}

When all of these criteria are met, the resulting signal $x_0(t)$ is said to be the first IMF for the input signal. When this resulting signal is subtracted from the original data, the residual can be taken as the new data, enabling the process to start again. If the criteria are met after following the steps and iterating upon them, a second order IMF, $IMF_1(t)$ may be defined. This process is then repeated, extracting new IMFs defined as $IMF_n(t)$, up to the point where the residual becomes a monotonic function (the trend of $x(t)$). 

\clearpage 

\section{Algorithm accuracy under ideal conditions}
\label{App:algo}
\begin{figure}[!t]
    \begin{center}
        \includegraphics[width=\textwidth]{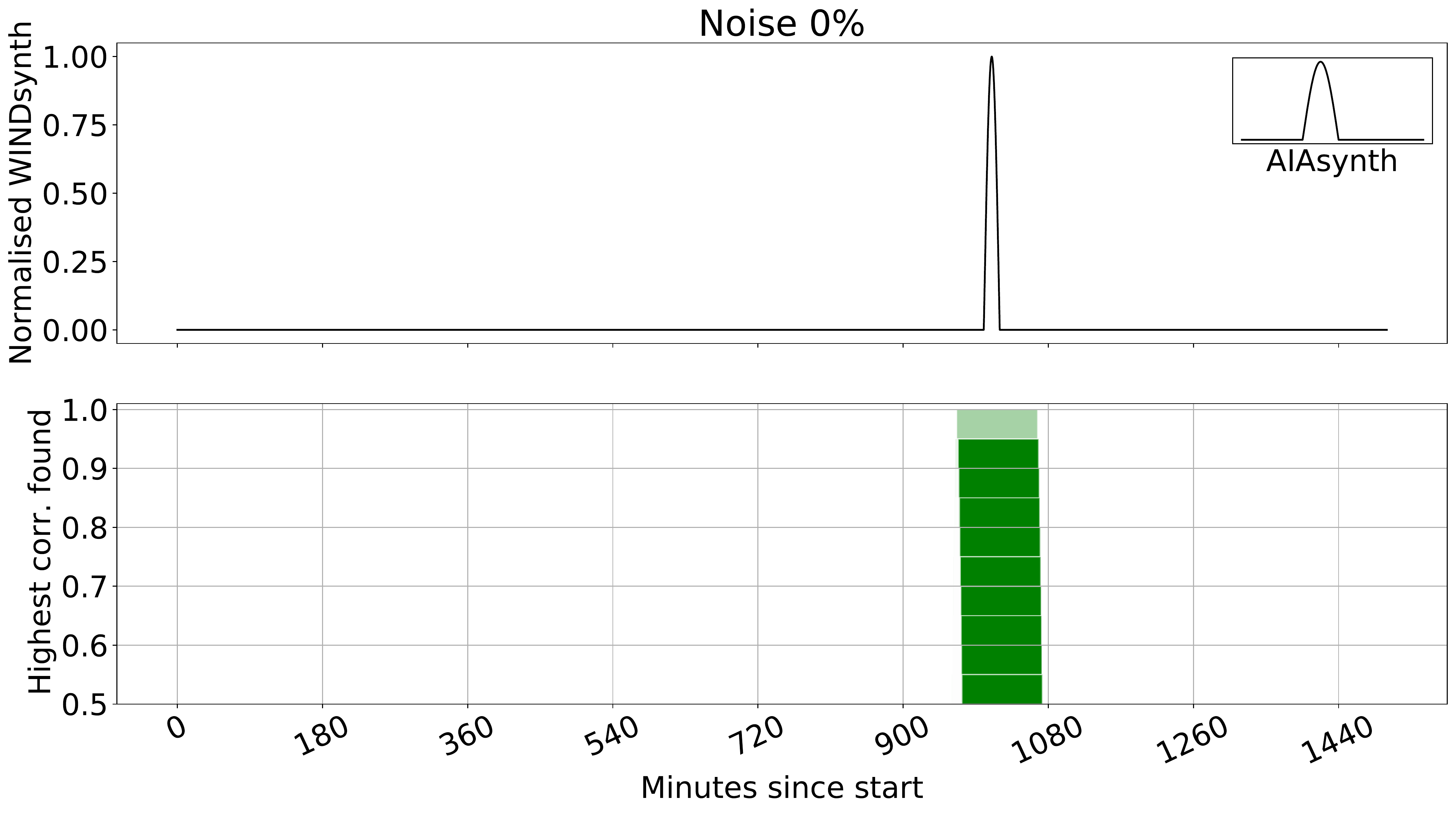}
    \end{center}
    
    \caption{Figures displaying locations where the correlation between the IMFs of AIA$_{synth}$ and WIND$_{synth}$ display high correlation for signals with no additional noise. On the top right, the AIA${synth}$ signal is shown. The X-axis is time since start in minutes, and the the Y-axis is the correlation threshold reached. Colour of the bars indicates how many IMFs display a correlation above the given level, with green for one pair, blue for two, and red for three or more IMF pairs.}
    \label{F-noiseless}
\end{figure}

In order to determine reliability of the algorithm, we perform a simple test where we embed a single Gaussian on a long, and a short dataset. We apply the smallest possible window displacement, and expect there to be several windows around the location of the embedded signal where the match is optimal, generating 100\% hit-rate as the rest of the signal does not correlate strongly.

The top panel of Figure \ref{F-noiseless} shows the WIND$_{synth}$ dataset, without any additional noise. On the top right, we show AIA$_{synth}$, a simple Gaussian pulse, whose IMFs we correlate against the normalised WIND$_{synth}$ signal. The bottom panel of Figure \ref{F-noiseless} displays correlation reached over time, specified in minutes since start, with the width of the bars being equal to the time duration of AIA$_{synth}$, and the green colour declaring that one IMF pair reaches each of the relevant correlation thresholds.

As we can see in Figure \ref{F-noiseless}, the only time where matching signatures are found within the derived IMFs is when the two signals overlap exactly, illustrated by the centre of the green bar coinciding with the peak of the Gaussian. Note that the colour of the match is slightly lighter as there is exactly one timestep where the two timeseries overlap exactly, and thus a correlation of exactly 1 is achieved. This test shows that our algorithm successfully finds the similarity between the two datasets, with 100\% hit-rate.

\clearpage

\section{\C{Significance of Correlation values for WIND case study}}
\label{App:pval}
\C{
In order to verify the statistical significance of the calculated correlations, we extracted the two-tailed p-values for the correlation matrix shown in Figure \ref{F-IMFcorrMatrixREAL}. The two-tailed p-value gives the probability that the null hypothesis can not be rejected. In the case of the Pearson R correlation, the null hypothesis is that the absolute value of a correlation value could be achieved by a different pair of datasets, therefore the p-value is a measure of the statistical significance of any calculated correlation. In this work, we take a p-value below 0.05 to be showing that the found correlation is statistically significant.}

\begin{figure}[!t]
    \begin{center}
        \includegraphics[width=0.72\textwidth]{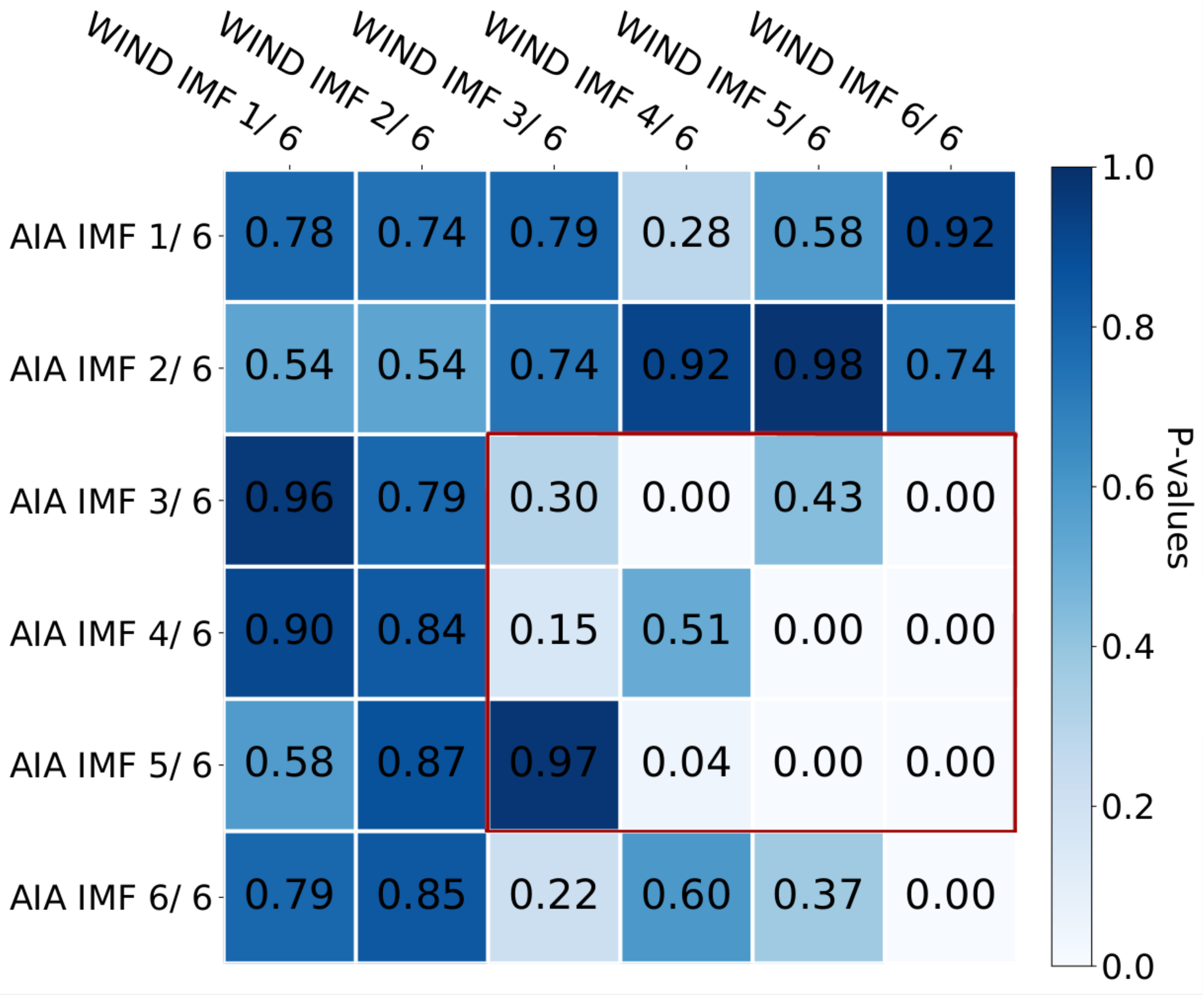}
    \end{center}
    
    \caption{\C{Matrix containing two-tailed p-values for IMF pairs calculated with the mass flux observational test case and the jetting signal shown in Section \ref{S-CStudy_WIND}. Intensity of the blue colouring is linked to p-value. Red highlight denotes IMF pairs which are used in analysis.}}
    \label{F-pvals}
\end{figure}

\C{We show p-values for the WIND case study shown in Section in Figure \ref{F-pvals}, with lighter colours referring to lower p-values, and darker colours referring to higher p-values. We note that in pairs of strong or moderate correlation within Figure \ref{F-AIA_real} (e.g., AIA IMF 5 : WIND IMF 6, with a Pearson R correlation of 0.9, or AIA IMF 4: WIND IMF 5, with a Pearson R correlation of 0.35), the p-value is low, and therefore the calculated Pearson R correlation is statistically significant. As the number of measurements is equal on all other parameters and correlations shown in Figure \ref{F-AIA_real}, as well as throughout the synthetic datasets from Section \ref{S-Algorithm-testing}, we expect moderate and strong correlation values (above 0.3) to be statistically significant throughout this work.}

\bibliographystyle{apalike}
\bibliography{main}

\end{article} 
\end{document}